\documentclass{article}

\usepackage{arxiv}
\usepackage[table,xcdraw,dvipsnames]{xcolor}
\usepackage[utf8]{inputenc} 
\usepackage[T1]{fontenc}    
\usepackage{hyperref}       
\usepackage{url}            
\usepackage{booktabs}       
\usepackage{amsfonts}       
\usepackage{nicefrac}       
\usepackage{microtype}      
\usepackage{lipsum}		
\usepackage{graphicx}
\usepackage{natbib}
\usepackage{doi}
\usepackage{tikz}

\usepackage{framed}
\usepackage{amsmath, amssymb, bm}
\usepackage{hyperref}
\usepackage{ragged2e}
\usepackage{booktabs, subfig}
\usepackage{pifont}
\usepackage{enumitem}
\usepackage{listings} 
\newcommand{\cmark}{\ding{51}}%
\newcommand{\xmark}{\ding{55}}%

\definecolor{FFFFFF}{HTML}{FFFFFF}
\definecolor{7987D7}{HTML}{7987D7}
\definecolor{A9E5BB}{HTML}{A9E5BB}
\definecolor{D6E3AE}{HTML}{D6E3AE}
\definecolor{E79C9C}{HTML}{E79C9C}
\definecolor{C5D86D}{HTML}{C5D86D}
\definecolor{D6A56F}{HTML}{D6A56F}
\definecolor{D7ADCD}{HTML}{D7ADCD}
\definecolor{FE5F55}{HTML}{FE5F55}
\definecolor{FFAA00}{HTML}{FFAA00}
\definecolor{388697}{HTML}{388697}
\definecolor{79D391}{HTML}{79D391}
\definecolor{8F52E0}{HTML}{8F52E0}
\definecolor{92DCE5}{HTML}{92DCE5}
\definecolor{CCD3D2}{HTML}{CCD3D2}
\definecolor{D884D2}{HTML}{D884D2}
\definecolor{F2DC5D}{HTML}{F2DC5D}

\title{stCEG: An R Package for Modelling Events over Spatial Areas Using Chain Event Graphs}

\author{ \href{https://orcid.org/0009-0002-1814-8428}{\includegraphics[scale=0.06]{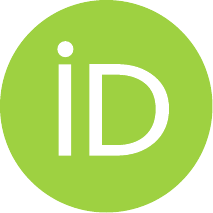}\hspace{1mm}Hollie Calley}\thanks{Corresponding author.} \\
    Department of Mathematics and Statistics\\
    University of Exeter\\
    \texttt{hc629@exeter.ac.uk} \\
	\And
	\href{https://orcid.org/0000-0001-8917-3300}{\includegraphics[scale=0.06]{orcid.pdf}\hspace{1mm}Daniel Williamson} \\
	Land, Environment, Economics and Policy Institute\\
    University of Exeter
}

\renewcommand{\shorttitle}{stCEG: Modelling Events over Spatial Areas Using CEGs}

\hypersetup{
pdftitle={stCEG: Modelling Events over Spatial Areas Using CEGs},
pdfsubject={Chain Event Graphs},
pdfauthor={Hollie Calley, Daniel Williamson},
pdfkeywords={chain event graphs, staged trees, Shiny, R, decision support},
}

\begin{document}
\maketitle

\begin{abstract}
	\textbf{stCEG} is an R package which allows a user to fully specify a Chain Event Graph (CEG) model from data and to produce interactive plots. It includes functions for the user to visualise spatial variables they wish to include in the model. There is also a web-based graphical user interface (GUI) provided, increasing ease of use for those without knowledge of R. We demonstrate \textbf{stCEG} using a dataset of homicides in London, which is included in the package. \textbf{stCEG} is the first software package for CEGs that allows for full model customisation.
\end{abstract}

\keywords{Chain Event Graphs \and Staged Trees \and \textbf{Shiny} \and R \and Decision Support}

\section[Introduction]{Introduction} \label{sec:intro}
There are many domains in which researchers use probabilistic graphical models (PGMs) to help inform decisions. PGMs combine both probability and graph theory, with the result being a graphical representation of the conditional independence relationships between events/variables in a system along with its underlying statistical model. A key benefit of this type of model is that the graph can often be understood regardless of the level of statistical training of the reader, making PGMs extremely useful for facilitating communication between statisticians, those who have domain knowledge, and decision makers (\cite{Walley2023Cegpy:Python}).

Bayesian Networks (BNs), first introduced in \cite{Pearl1986FusionNetworks} are by far the most popular family of PGM, used across a variety of domains such as crime, medicine and agriculture. They model processes of events, sequential unfoldings of choices and outcomes, where each decision point is influenced by prior events and leads to a range of possible future paths. 
However, they exhibit some limitations which restrict the types of processes that can be fully modelled using them, namely their inability to model asymmetry. There are two main types of asymmetry we consider in this paper: 
\begin{itemize}
    \item Asymmetric Conditional Independence: A process can be asymmetric in its conditional independence relationships; there are independences that only hold for certain values of the variable it is conditioned on. For example, given variables $A,B,C$, we may have $C \perp\!\!\!\perp B \mid A$ for some values of $A$, but $C \not\!\perp\!\!\!\perp B \mid A$ for others.
    \item Asymmetric Structure: These are processes containing missing values which logically cannot be non-zero. For example, considering we are scheduling people for a series of surgeries, we cannot logically schedule for a second surgery if they did not survive the first.
\end{itemize}

When explaining processes of events, experts in many domains such as policing \cite{Boutilier1996Context-SpecificNetworks}, public health \cite{Shenvi2019ModellingGraphs} and risk analysis \cite{Pensar2015LabeledModels} often describe these as asymmetric systems (with one or both types of asymmetry), yet BNs are unable to fully model these.

Chain Event Graphs (CEGs) were developed by \cite{Smith2008ConditionalGraphs} as an alternative to the BN that can handle asymmetry. They arise from a tree structure, partitioning the variables in a dataset so that each logical combination of variables follows a unique path through the tree. This is incredibly intuitive when describing a sequence of events, as it is clear to see which events lead to which outcomes. Vertices in these trees are coloured according to their conditional independence structure, with identically coloured nodes representing equivalent independence statements. The graph structure is then reduced on the basis of this colouring, giving a graph with the minimum number of nodes and edges. A technical review of this process can be found in Section 2. Since their initial development, various theoretical developments have been made to the CEG, such as model selection (\cite{Freeman2011BayesianGraphs}), equivalence classes (\cite{Gorgen2018EquivalenceTrees}), d-separation (\cite{Wilkerson2020CustomisingModels}), and dynamic variants (\cite{Shenvi2021Non-StratifiedApplications, Freeman2010LearningGraphs, Barclay2015TheGraph, Collazo2018AnGraph}). However, while CEGs have been extended to model both continuous and discrete time, research has not yet focused on extending them to model spatial events.  

Although CEGs have been proven to be more flexible than BNs in terms of the situations that they can model, their use remains limited. It has been suggested that the primary reason for this is the lack of existing software (\cite{Strong2023MethodologicalGraphs}).Currently there are 3 software packages that exist for modelling using CEGs, \textbf{stagedtrees} (\cite{Carli2022TheTrees}) and \textbf{ceg} (\cite{Collazo2017Ceg:Graph}) in R, and \textbf{cegpy} (\cite{Walley2023Cegpy:Python}) in Python. 

\vspace{1em}
\begin{table}[ht!]
\centering
\begin{tabular}{l|l|l|l|l}
\hline
                           & \textbf{cegpy}      & \textbf{stagedtrees} & \textbf{ceg}    & \textbf{stCEG}  \\ \hline
Language                   & Python & R      & R & R \\ 
Support for asymmetry        & \cmark           & \cmark            & \xmark       & \cmark       \\ 
Fully customised colouring & \xmark           & \cmark            & \xmark       & \cmark       \\ 
Fully customised priors    & \cmark           & \xmark            & \xmark       & \cmark       \\ 
Support for spatial variables    & \xmark           & \xmark            & \xmark       & \cmark       \\
Interactive/User Interface & \xmark           & \xmark            & \xmark       & \cmark       \\ 
\end{tabular}
\vspace{1em}
\caption{Comparison of the packages currently available for modelling with CEGs}
\label{table:packagecomparisons}
\end{table}

Table \ref{table:packagecomparisons} gives an overview of the key features of each package. Both \textbf{cegpy} and \textbf{stagedtrees} are now able to model processes with structural asymmetry (when initially published, \textbf{stagedtrees} did not have this functionality). Yet, despite the flexibility of CEGs, none of these packages allow the user to fully specify a model, that is to both colour the vertices fully and also specify their probability distributions. Furthermore, all of the packages require users to be experienced coders in order to engage fully with them. With a model type that aims to be interpretable for those without mathematical knowledge, this is an additional barrier for use. In this paper, we present \textbf{stCEG}, the first package for CEGs that allows full model customisation and is designed to incorporate spatial variables. It features both a set of functions that can be run in R, and an intuitive user interface in \textbf{Shiny} (\cite{Chang2024Shiny:R}), so that users without coding experience can still access the package. We will walk through both the functionality of the package itself and of the user interface.

\section{Staged Trees and Chain Event Graphs} \label{sec:stagedtreesandcegs}

Chain Event Graphs (CEGs) are formed by first constructing an event tree, which describes a process of events, with each combination of logical variables having a path through the tree. This structure may be symmetric or asymmetric. Asymmetry is represented in an event tree by the omission of edges which are logically impossible, reducing the size of the model structure (as no unnecessary information is included). We illustrate the difference between symmetric and asymmetric event tree structures in Figure \ref{EventTreesSymmetry} below.

\begin{figure}[ht!]
    \centering
    \subfloat[Symmetric event tree representing possible effects of weapon and gender on the solved rate of homicide cases]{%
        \includegraphics[width=0.41\textwidth]{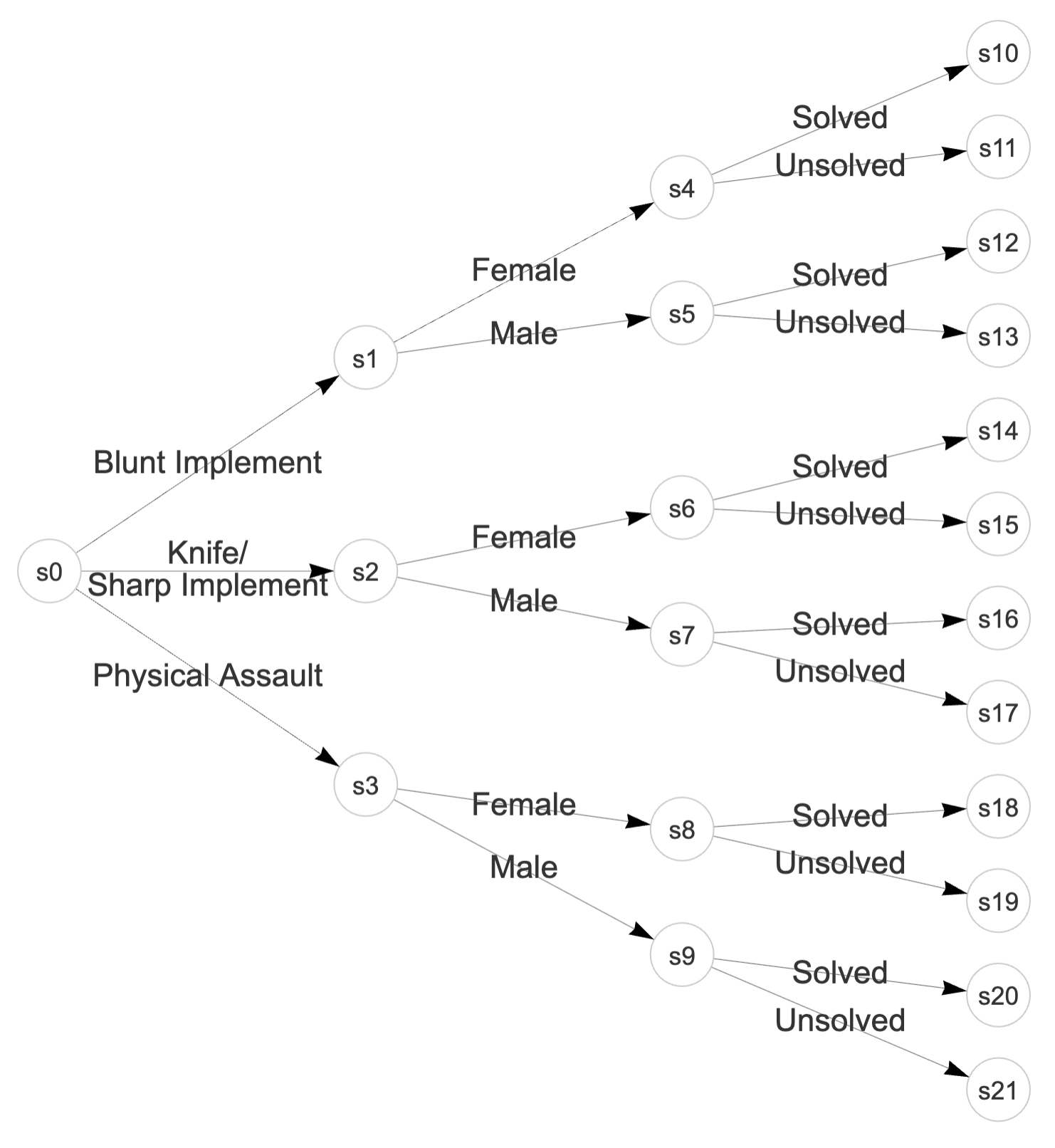}
        \label{fig:1a}
    }
    \hfill
    \subfloat[Asymmetric event tree representing how evidence types affect the likelihood of a suspect being charged.]{%
        \raisebox{12mm}{\includegraphics[width=0.46\textwidth]{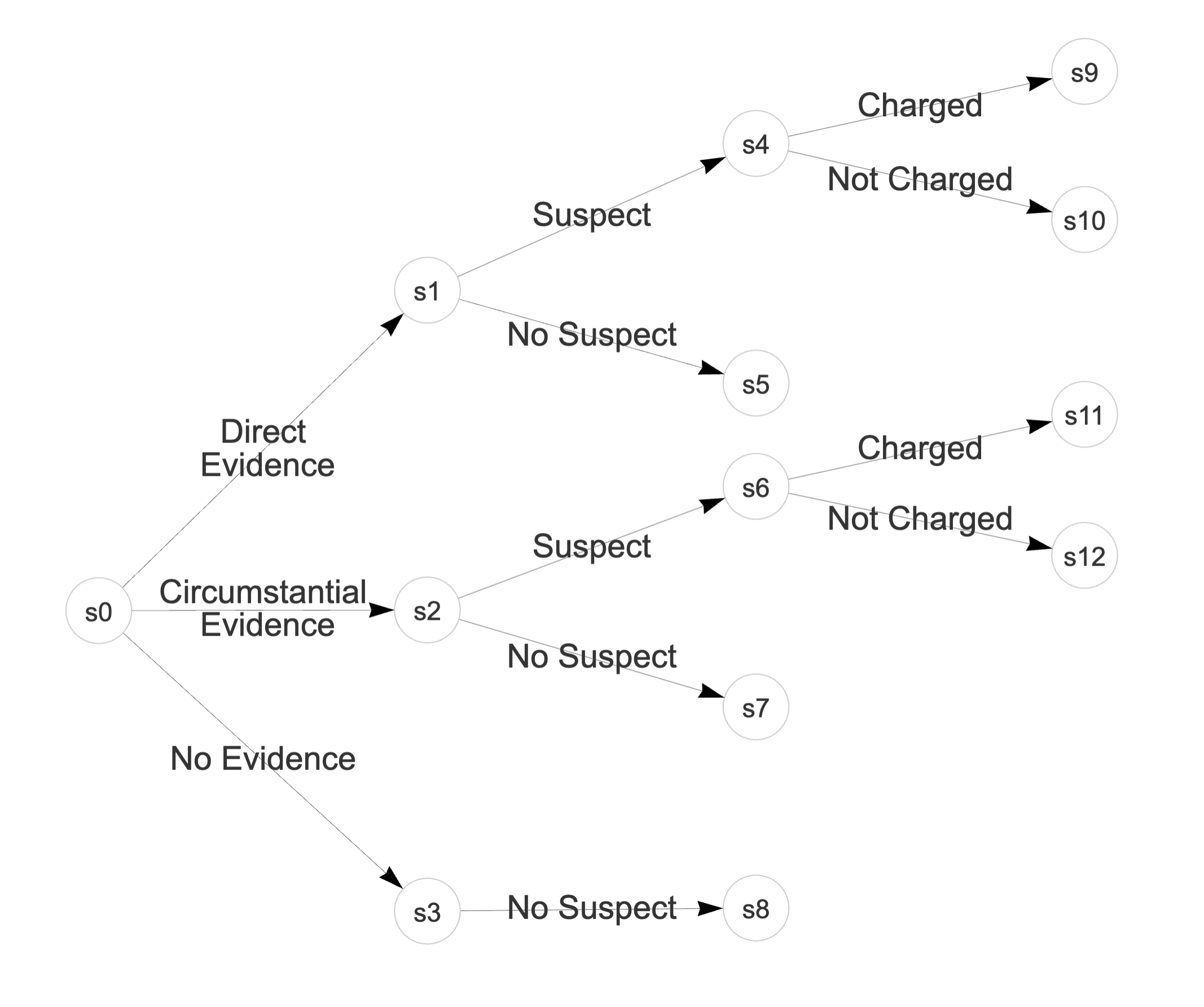}}
        \label{fig:1b}
    }
    \caption{Comparison of symmetric and asymmetric event trees for two different crime scenarios.}
    \label{EventTreesSymmetry}
\end{figure}

\begin{figure}[h!]
    \centering
    \subfloat[Event tree with \textcolor{YellowOrange}{root node}, \textcolor{RoyalBlue}{situations}, and \textcolor{RedOrange}{leaf nodes}.]{%
        \includegraphics[width=0.44\textwidth]{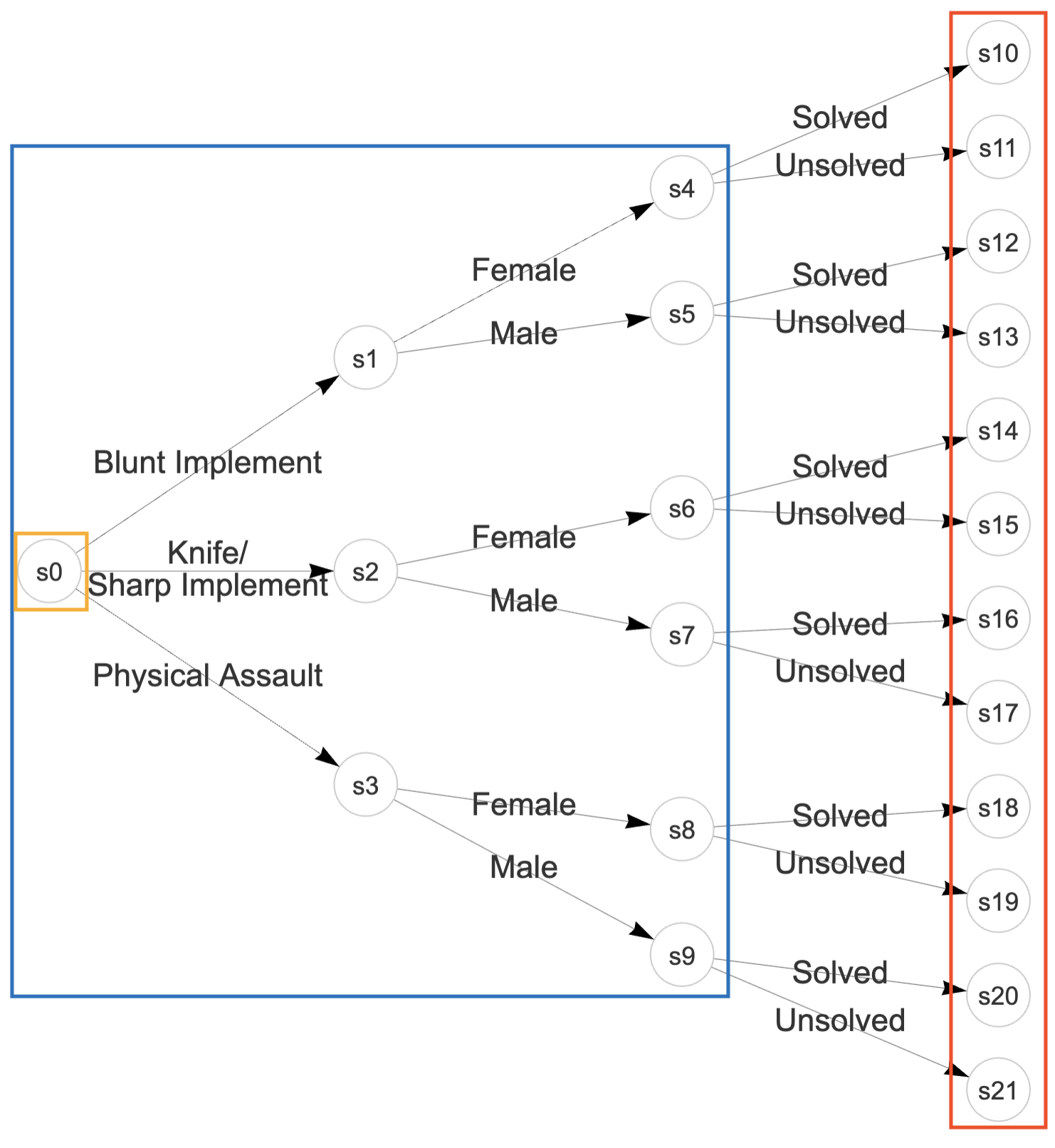}
            
    \label{fig:highlightedET}
    }
    \hfill
    \subfloat[Staged tree representing one possible set of exchangeability judgements.]{
\includegraphics[width=0.44\textwidth]{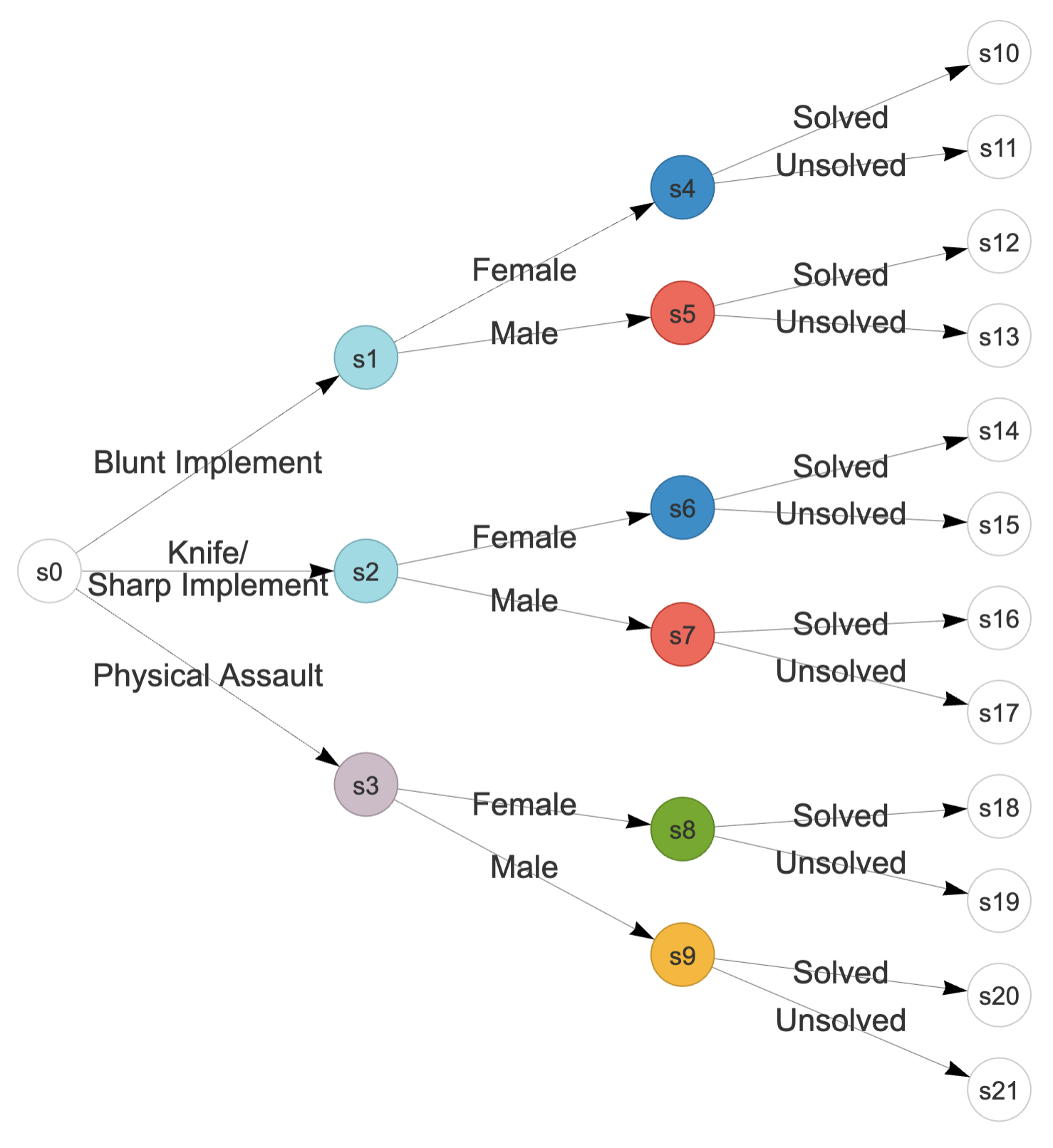}
        \label{fig:SymmetricST}
    }
    \caption{Event tree $\mathcal{T}$ with features highlighted and corresponding staged tree $\mathcal{S}$, showing one possible staging $U$.}
    \label{EventTreesStaging}
\end{figure}

An event tree can be formally defined as $\mathcal{T} = (V(\mathcal{T}), E(\mathcal{T}))$, where $V(\mathcal{T})$ represents the set of vertices and $E(\mathcal{T})$ the set of directed edges. The vertex $s_0$ is referred to as the root node. Vertices with at least one outgoing edge are called situations, $S(\mathcal{T})$, while those that appear at the end of a path in the tree are known as leaf nodes, $L(\mathcal{T})$. Figure \ref{fig:highlightedET} indicates these features on the event tree in Figure \ref{fig:1a}. 

The collection of all root-to-leaf paths in $\mathcal{T}$ forms the event space, denoted $\mathbb{X} = \{\lambda(s_0, s) : s \in L(\mathcal{T})\}$. For each situation $s \in S(\mathcal{T})$, there is an associated random variable $X(s)$, defined conditionally upon reaching $s$. The corresponding state space is written as $\mathbb{X}(s)$.
Given an event tree, $\mathcal{T}$, exchangeability judgments are encoded through vertex colouring, where vertices assigned the same colour are assumed to share identical probability distributions. These groups of vertices are referred to as stages, and the resulting coloured tree is known as a staged tree, denoted $\mathcal{S}$, with staging $U$. An example of a staged tree can be found in Figure \ref{fig:SymmetricST}.

Formally, two vertices $s_i, s_j \in \mathcal{T}$ belong to the same stage $u$ if and only if there exists a bijection

\begin{equation*}
  \psi_u(s_i, s_j): \mathbb{X}(s_i) \rightarrow \mathbb{X}(s_j)  
\end{equation*}

\vspace{0.5em}

under which the random variables $X(s_i)$ and $X(s_j)$ follow the same probability distribution. This reflects the belief that the distributions for the outcomes immediately following $s_i$ and $s_j$ are identical (the arrows directly succeeding $s_i$ and $s_j$). In Figure \ref{fig:SymmetricST}, the shared colouring of $\{s_1, s_2\}$ implies that if the method of killing is either ``Blunt Implement'' or ``Knife/Sharp Implement'', then the probability distribution over the victim’s sex (``Female'' or ``Male'') is assumed to be the same in both cases.

Vertex colouring is performed, with a probability distribution then assigned to each stage. This colouring can either be done using some combination of expert judgement and algorithms such as Agglomerative Hierarchical Clustering (AHC), first implemented by \cite{Freeman2011BayesianGraphs}. The AHC algorithm starts from a fixed variable ordering, with all vertices being in their own stage. It then merges pairs of vertices to maximise the log marginal likelihood score. Technical details of the AHC algorithm can be found in Appendix \ref{sec:ahcalgorithm}, and furhter discussed in \cite{Collazo2018ChainGraphs}. Other Bayesian algorithms have been explored for CEGs that search over all possible variable orderings, such as the dynamic programming methods discussed in \cite{Silander2013AGraphs} and \cite{Cowell2014CausalGraphs}, however due to the computational power needed to search over all possible models, they are currently beyond the scope of \textbf{stCEG}.

Given the structure of $\mathcal{S}$, we can now specify prior judgements for each of our stages. We have coloured $\mathcal{S}$ with $k$ unique colours corresponding to stages $u_1,\ldots,u_k$, and each vertex in stage $u_i$ has $k_i$ edges leaving it. Each stage $u_i$ with $k_i$ outgoing edges has a conditional probability vector (CPV). This gives the probability that an individual at some $s \in u_i$ travels along its \textit{j}th outgoing edge, where $j \in \{1,\ldots,k_i\}$. This can be written for each stage as $\bm{\theta}_i = (\theta_{i1}, \theta_{i2},\ldots,\theta_{ik_i})$.

We can specify these initial CPVs by assigning parameters for a prior probability distribution. \cite{Freeman2010LearningGraphs} argues for the use of the Dirichlet distribution for prior specification of CEGs due to Dirichlet-Multinomial conjugacy, so that our posterior is analytically tractable. We then assume that our CPV $\bm{\theta}_i$ for stage $u_i$ with $k_i$ outgoing edges has a Dirichlet prior distribution $\text{Dir}(\bm{\alpha}_i) = \text{Dir}(\alpha_{i1},\ldots,\alpha_{ik_i})$. This also enables the user to express their uncertainty through the magnitude of the specified $\alpha_{ij}$ parameters.

\vspace{0.5em}

\begin{table}[h!]
    \centering
\begin{tabular}{|l|l|l|l|l|}
    \hline
    Stage & Situations & \begin{tabular}[c]{@{}l@{}}Outgoing edges per\\ vertex\end{tabular} & \begin{tabular}[c]{@{}l@{}}Prior parameters\end{tabular} & \begin{tabular}[c]{@{}l@{}}Conditional probability \\ vector (CPV)\end{tabular} \\ \hline
    $u_1$ & $\{s_0\}$ & 3 & (46,69,115) & (0.2, 0.3, 0.5) \\ \hline
    \cellcolor[HTML]{92DCE5}$u_2$ & $\{s_1\}, \{s_2\}$ & 2 & (27,73) & (0.27, 0.73) \\ \hline
    \cellcolor[HTML]{CEBBC9}$u_3$ & $\{s_3\}$ & 2 & (31,19) & (0.62, 0.38) \\ \hline
    \cellcolor[HTML]{008FCC}$u_4$ & $\{s_4\}, \{s_6\}$ & 2 & (17,3) & (0.85, 0.15) \\ \hline
    \cellcolor[HTML]{FE5F55}$u_5$ & $\{s_5\}, \{s_7\}$ & 2 & (49,51) & (0.49, 0.51) \\ \hline
    \cellcolor[HTML]{66AA00}$u_6$ & $\{s_8\}$ & 2 & (23,2) & (0.92, 0.08) \\ \hline
    \cellcolor[HTML]{FFB400}$u_7$ & $\{s_9\}$ & 2 & (17,8) & (0.68, 0.32) \\ \hline
\end{tabular}
\vspace{1em}
\caption{Situation set, number of outgoing edges, prior parameters, and CPV for each stage in $\mathcal{S}$}
\label{tab:CPVs}
\end{table}

Whilst experts who have mathematical knowledge may be able to specify a suitable prior distribution for the model, different methods have been proposed for when we wish to utilise knowledge from domain experts regardless of their mathematical background. \cite{Collazo2018AnGraph} proposed that due to Dirichlet-Multinomial conjugacy, that the prior could be viewed as the number of phantom units apriori that the expert believes will traverse each edge. The potential pitfall of this is that it may appear to the expert that the phantom samples must sum across the CEG. This could cause an expert to make overconfident judgements on a branch of the CEG they are unsure about due to this false summing constraint. To this end, \cite{Shenvi2021Non-StratifiedApplications} proposed an approach in which experts are asked two questions in order to form a prior distribution over each stage:  
\begin{enumerate}
    \item How many people would they be comfortable sending down a set branch of the CEG (e.g. ``Physical Assault'').
    
    If the expert is uncertain, this may be a small number. If they are very certain, this number may be much larger. The larger the number the expert assigns, the more weight this value will be given compared to the observed data values. It is beneficial for the expert to know roughly how many individuals are in the system, so they can judge the strength of their chosen prior.
    \item Given the number of people the expert chose for Question 1, how would the expert choose to split these into the next category (e.g. ``Female'' or ``Male'')?
    
This enables us to specify Dirichlet prior parameters for a given vertex. For example, if an individual stated that they would assign 100 individuals to the 'Physical Assault' branch of the staged tree in Figure \ref{fig:SymmetricST}, anticipating that 62 would subsequently follow the ``Female'' branch and 38 the ``Male'' branch, the corresponding Dirichlet prior for that stage would be $\text{Dir}(62, 38)$.

\end{enumerate}

It is useful to note the difference between setting priors on the situations and setting priors on the stages. Due to the priors being Dirichlet, if the priors for each of the situations in the stage are specified separately, the stage prior  is the sum of these over each edge label, namely $\sum_{j=1}^{k_i}\alpha_{ij}$ (\cite{Freeman2011BayesianGraphs}).
If the expert is unsure either about the system as a whole or a subset of situations in the system, one method of accounting for this is to set an uninformative Dirichlet distribution for those stages. This is usually either a Dir$(1,\ldots,1)$ distribution for each situation, or a ``Phantom Individuals'' prior, as described in \cite{Collazo2018AnGraph} and implemented in \cite{Walley2023Cegpy:Python}.Our initial sample size, $\alpha$ is set as the maximum number of branches coming from any vertex in $\mathcal{S}$, and this starts at $s_0$ and is evenly divided throughout the tree. 

Given we have specified prior distributions for each stage $u_i \in \mathcal{S}$, we can perform an update on the system. This requires a data vector $\bm{y}_i = (y_{i1}, y_{i2},\ldots, y_{ik_i})$ of individuals passing through each stage $u_i$, modelled as Multinomial. Using Bayes Theorem, our posterior for stage $u_i$ is then also a Dirichlet distribution, with parameters $\text{Dir}(\alpha_{i1} + y_{i1},\ldots,\alpha_{ik_i}+ y_{ik_i})$. The derivation of this can be found in Appendix \ref{app:dirichletcalculation}.
\vspace{1em}
\begin{table}[ht!]
\centering
\begin{tabular}{|l|l|l|l|l|l|}
\hline
Stage                         & Prior       & Prior Mean      & Data $(\bm{y}_i)$ & Posterior     & Posterior Mean    \\ \hline
$u_1$                         & (46,69,115) & (0.2, 0.3, 0.5) & (68,86,245)       & (114,155,360) & (0.18,0.25, 0.57) \\ \hline
\cellcolor[HTML]{92DCE5}$u_2$ & (27,73)     & (0.27, 0.73)    & (67,87)           & (94,160)      & (0.37,0.63)       \\ \hline
\cellcolor[HTML]{CEBBC9}$u_3$ & (31,19)     & (0.62, 0.38)    & (98,147)          & (129,166)     & (0.44,0.56)       \\ \hline
\cellcolor[HTML]{008FCC}$u_4$ & (17,3)      & (0.85, 0.15)    & (61,6)            & (68,9)        & (0.88,0.12)       \\ \hline
\cellcolor[HTML]{FE5F55}$u_5$ & (49,51)     & (0.49, 0.51)    & (64,23)           & (113,74)      & (0.60,0.40)       \\ \hline
\cellcolor[HTML]{66AA00}$u_6$ & (23,2)      & (0.92, 0.08)    & (78,20)           & (101,22)      & (0.82,0.18)       \\ \hline
\cellcolor[HTML]{FFB400}$u_7$ & (17,8)      & (0.68, 0.32)    & (112,35)          & (129,43)      & (0.75, 0.25)      \\ \hline
\end{tabular}
\vspace{1em}
\captionof{table}{Situation set, number of outgoing edges and prior CPV for each stage in $\mathcal{S}$}
        \label{tab:update}
\end{table}

This update can be represented on the staged tree $\mathcal{S}$, however when increasing the number of variables these structures often become large and hard to read from. 

\begin{figure*}[ht!]
    \centering      \includegraphics[width=0.63\textwidth]{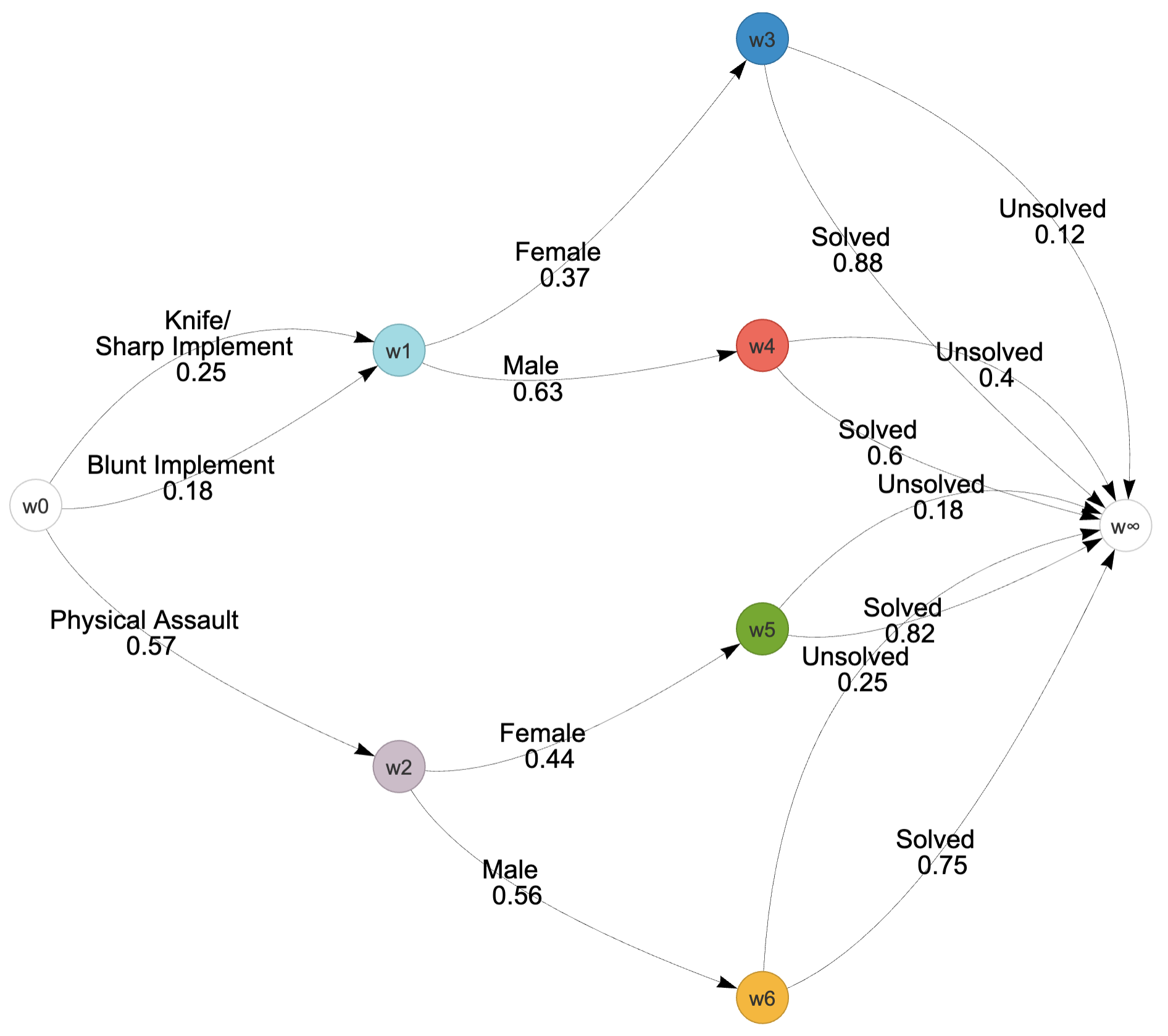}
    \caption{Chain Event Graph obtained from the corresponding staged tree in Figure \ref{fig:SymmetricST}, with posterior mean values on each edge}
    \label{fig:SymmetricCEG}
\end{figure*}

We can convert $\mathcal{S}$ into a Chain Event Graph $\mathcal{C}$ by merging the vertices whose colour and structure are the same. Two vertices, $s_i$ and $s_j$, in $\mathcal{S}$ are said to have the same structure if and only if the staged trees rooted at $s_i$ and $s_j$ respectively, also known as florets, have the same colouring. All leaf vertices $L(T)$ are contracted into a single vertex $w_{\infty}$, known as the sink vertex. From Figure \ref{fig:SymmetricST}, we can see that the florets rooted at $s_1$ and $s_2$ are identical, hence they are merged into vertex $w_1$ in Figure \ref{fig:SymmetricCEG}. This reduces the size of the graph without any loss of information, making it easier to infer relationships between variables.

\section{The stCEG Package} \label{sec:stceg}

We now explore the functionality of the \textbf{stCEG} package: 
\subsection{Creating an Event Tree}
\label{subsec:createET}
\textbf{stCEG} has 3 main classes: \texttt{event$\_$tree}, \texttt{staged$\_$tree} and \texttt{chain$\_$event$\_$graph}, which all make use of the \textbf{visNetwork} (\cite{Almende2025VisNetwork:Library}) package for producing interactive plots. The function \texttt{create$\_$event$\_$tree} creates an output of class \texttt{event$\_$tree}. It requires an input dataset (in \texttt{data.frame} format) to form a symmetric tree using all combinations of selected factors. The user can then choose any nodes to be deleted using the function \texttt{delete$\_$nodes} if the system requires an asymmetric structure. Paths are ordered alphabetically, so node deletion does not affect the labels of surrounding nodes. This differs from the method implemented in \textbf{cegpy}, which maps only observed paths from the dataframe, with the user then having to specify any unobserved edges individually. In cases where there are many logically plausible branches with unobserved values, this is often incredibly time consuming. The output is comprised of two objects: \texttt{eventtree}, a \textbf{visNetwork} object displaying the tree, along with \texttt{filtereddf}, the dataframe that created it. The \textbf{visNetwork} plot allows the user to click and drag nodes, giving freedom to rearrange nodes as desired.

\subsection{Colouring an Event Tree for conversion into a Staged Tree}\label{subsec:stcegcolouring}
Unlike other software packages for CEGs, \textbf{stCEG} allows for a completely custom colouring, using a method of both expert judgement and colouring using Agglomerative Hierarchical Clustering (described in Section \ref{sec:stagedtreesandcegs}, and further detailed in Appendix \ref{sec:ahcalgorithm}).This is done using the functions \texttt{update$\_$node$\_$colours} to manually specify stages, and \texttt{ahc$\_$colouring} to use the AHC algorithm, outputting objects with class \texttt{staged$\_$tree}. Both of these functions can be used in either order. The user can colour a subset of vertices to express any judgements they may have whilst being able to remain uncertain for vertices if desired, preventing overconfidence. The AHC algorithm can then be applied to any vertices that have not been manually coloured. The AHC can also be initially applied to the whole tree, and then the user can edit groups as desired.
When specifying stages, the user has the ability to assign specific colours using the \texttt{colours} argument in \texttt{update$\_$node$\_$colours}, allowing for further customisation of the \textbf{visNetwork} output. This is discussed further with an example in Section \ref{subsec:colouringtree}. The output from both of these colouring functions is a \texttt{stagedtrees} \textbf{visNetwork}, along with the \texttt{filtereddf} as before, this time with class \texttt{staged$\_$tree}.

\subsection{Prior Specification}\label{subsec:PriorSpec}
\begin{figure*}[h!]
    \centering  
    \includegraphics[width=0.5\textwidth]{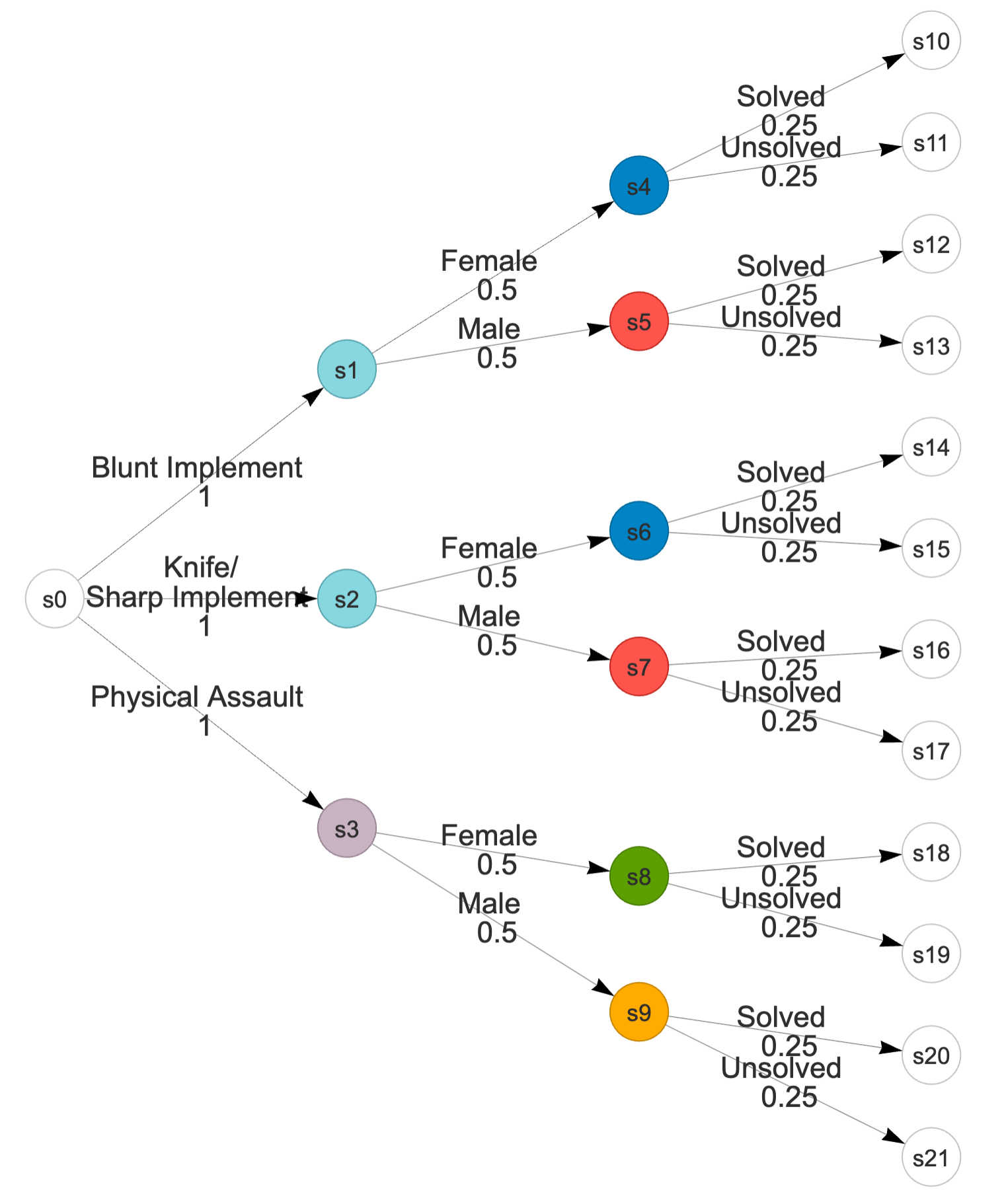}
    \caption{Staged Tree $\mathcal{S}$, with ``Phantom'' priors on each stage, as described in \cite{Collazo2018ChainGraphs}. $\alpha = 3$, as this is the maximum number of outgoing edges for any situation $s_i$.}
    \label{fig:PhantomStagedTree}
\end{figure*}

Given that the user has fully coloured the event tree, using either of the functions described above, they can then use the function \texttt{specify$\_$priors} in order to give each stage a Dirichlet prior. \textbf{stCEG} allows for 3 different prior types: 
\begin{itemize}

    \item \texttt{Custom}: This allows the user to manually specify the priors for each stage. 
    \vspace{0.5em}
    \item \texttt{Uniform}: This assigns a $\text{Dirichlet}(1,\ldots,1)$ prior, $\alpha_{ij} = 1$ for each outgoing edge from situation $s_i$ in stage $u_i$. 
    \vspace{0.5em}
    \item \texttt{Phantom}: Using the approach of \cite{Collazo2017Ceg:Graph}, this initialises an imaginary sample size, $\alpha$, the maximum number of outgoing edges from any situation $s_i$ in $\mathcal{S}$, and divides that evenly throughout the tree (see Figure \ref{fig:PhantomStagedTree}).
    
\end{itemize}

If the prior type is selected as either ``\texttt{Uniform}'' or ``\texttt{Phantom}'', then the user is given the option to edit priors for specific stages via a console prompt:

\vspace{1em} 

\texttt{Do you want to edit specific rows? (yes/no): \\
\textcolor{blue}{1: yes}\\
Enter row numbers to edit (comma-separated, e.g., 1,3,5): \\
\textcolor{blue}{1: 1,4,7}\\
Enter new prior for row 1 (3 values, comma-separated): \\
\textcolor{blue}{1: 34,25,12}\\
Enter new prior for row 4 (2 values, comma-separated): \\
\textcolor{blue}{1: 60,40}\\
Enter new prior for row 7 (2 values, comma-separated): \\
\textcolor{blue}{1: 76,8}}

\newpage 

This function (with or without edits) outputs a dataframe with the Dirichlet priors and corresponding prior CPVs for each stage, as in Table \ref{tab:CPVs}. The flexibility of being able to remain uncertain about situations in a event tree $\mathcal{T}$, both in terms of their staging and the prior distribution of their stages makes \textbf{stCEG} the most flexible package for CEGs in terms of its prior modelling capabilities.

The function \texttt{staged$\_$tree$\_$prior}, takes an output of \texttt{stagedtree} class from either \texttt{ahc$\_$colouring} or \texttt{update$\_$node$\_$colours}, along with the dataframe output from \texttt{specify$\_$priors}, and produces a \texttt{stagedtreewithpriors} \textbf{visNetwork} graph. The user can either choose to display priors, prior means, or only display the variable names, using the \texttt{label$\_$type} argument. An example \textbf{visNetwork} graph can be found in Figure \ref{fig:PhantomStagedTree}, with \texttt{label$\_$type = ``priors''}, and a worked example along with reproducible code can be found in Section \ref{subsec:specify_priors}. Hovering over a situation in the \textbf{visNetwork} output gives the prior distribution, mean and variance for that stage.

\subsection{Reducing a Staged Tree into a Chain Event Graph}

To convert a staged tree into a CEG using the process discussed in Section \ref{sec:stagedtreesandcegs}, the \texttt{create$\_$ceg} function is applied to an output from \texttt{staged$\_$tree$\_$prior}. Similarly to the above function, the user can choose to display either the posterior mean or posterior Dirichlet parameters for each situation via the \texttt{label} argument. 

To aid readability, especially in the case of large models, clicking on a situation in the CEG output highlights the edges entering it in blue, and those leaving it in red. Furthermore, setting \texttt{view$\_$table = TRUE} outputs a table showing the update parameters for each stage (similar to Table \ref{tab:update}). This table is shown in the viewer as an image instead of in the console, as R does not support printing colour codes in a table. The colour coded table is useful for tracking specific paths through the tree in the case that these are hard to manually follow in the CEG, and also to assess the strength of the priors for each stage by exploring the difference between the prior and posterior means. 

Where the model structure is large, the user may wish to filter the CEG to show only a subset of categories using the function \texttt{create$\_$reduced$\_$CEG}. This produces a plot of the paths in the CEG which include selected categories chosen with the argument \texttt{filter$\_$by = c("Var1", "Var2")}. This can aid in comparison between multiple categories, as smaller graphs with fewer variables are easier to read from.

We can also compare two models using the same data with the function \texttt{compare$\_$ceg$\_$models}. This calculates the log-marginal likelihood of both models, and uses this to form the log Bayes factor to give a quantifiable interpretation of support for one model over the other.

\subsection{Incorporating Spatial Characteristics into CEGs}
\textbf{stCEG} aims to increase the interpretability of CEG models which include space. The user is able to create a CEG containing spatial variables, and leverage these by linking it with a shapefile of the chosen area. The \texttt{generate$\_$CEG$\_$map} function is able to produce maps using the \textbf{leaflet} (\cite{Cheng2025Leaflet:Library}) package showing the outputs collated over each area, conditioned on user-chosen variables in the CEG by specifying these with the argument \texttt{conditionals = c("Var1", "Var2")}. 

\section{An illustrative example: Homicide data}
To illustrate the functionality of the \textbf{stCEG} package, we will use data sourced from the Metropolitan Police Homicide Dashboard (\cite{MetropolitanPolice2025HomicideDashboard}) to explore factors affecting the solved status of a homicide case. This data contains both demographic and crime characteristics for homicide victims between 2003-2023.
The subset of categorical variables used in this example are:
\begin{itemize}
\setlength{\itemsep}{0.5pt}
    \item Method of Killing: \{Blunt Implement, Knife or Sharp Implement, Physical Assault, Shooting\} \vspace{0.5em}
    \item Sex: \{Female, Male\} \vspace{0.5em}
    \item Domestic Abuse: \{Domestic Abuse, Not Domestic Abuse\}
    
    A homicide is classed as a ``domestic homicide'' if the suspected or charged individual is either related to, in a relationship with, or a member of the same household as the victim (\cite{HomeOffice2004Domestic2004}). \vspace{0.5em}
\item Solved Status: \{Solved, Unsolved\} 

Solved for crime-reporting purposes means either a charge has been made in the case, or a suspect has been identified, but for some reason a charge cannot be brought against them, for example if they are dead (\cite{HomeOffice2011UserStatistics}). \vspace{0.5em}
\end{itemize}

Loading the dataset into R, we can  view the first few rows:
\vspace{1em}

\texttt{\textcolor{blue}{library}(stCEG)}\\
\texttt{head(homicides)}

\vspace{0.5em}

\begin{table}[ht]
\centering
\begin{tabular}{rllll}

 & \texttt{Method\_of\_Killing} & \texttt{Sex} & \texttt{Domestic\_Abuse} & \texttt{Solved\_Status} \\ 
  
\texttt{1} & \texttt{Knife or Sharp Implement} & \texttt{Male} & \texttt{Not Domestic Abuse} & \texttt{Solved} \\ 
\texttt{2} & \texttt{Physical Assault, no weapon} & \texttt{Male} & \texttt{Not Domestic Abuse} & \texttt{Solved} \\ 
\texttt{3} & \texttt{Blunt Implement} & \texttt{Male} & \texttt{Not Domestic Abuse} & \texttt{Solved} \\ 
\texttt{4} & \texttt{Physical Assault, no weapon} & \texttt{Female} & \texttt{Not Domestic Abuse} & \texttt{Solved} \\ 
\texttt{5} & \texttt{Knife or Sharp Implement} & \texttt{Male} & \texttt{Not Domestic Abuse} & \texttt{Solved} \\ 
\texttt{6} & \texttt{Physical Assault, no weapon} & \texttt{Female} & \texttt{Not Domestic Abuse} & \texttt{Solved} \\ 
\end{tabular}
\vspace{-1em}
\end{table}

\vspace{0.5em}

\subsection{Forming the Event Tree}
\label{subsec:formET}

\begin{figure*}[ht!]
    \centering  
    \includegraphics[width=0.72\textwidth]{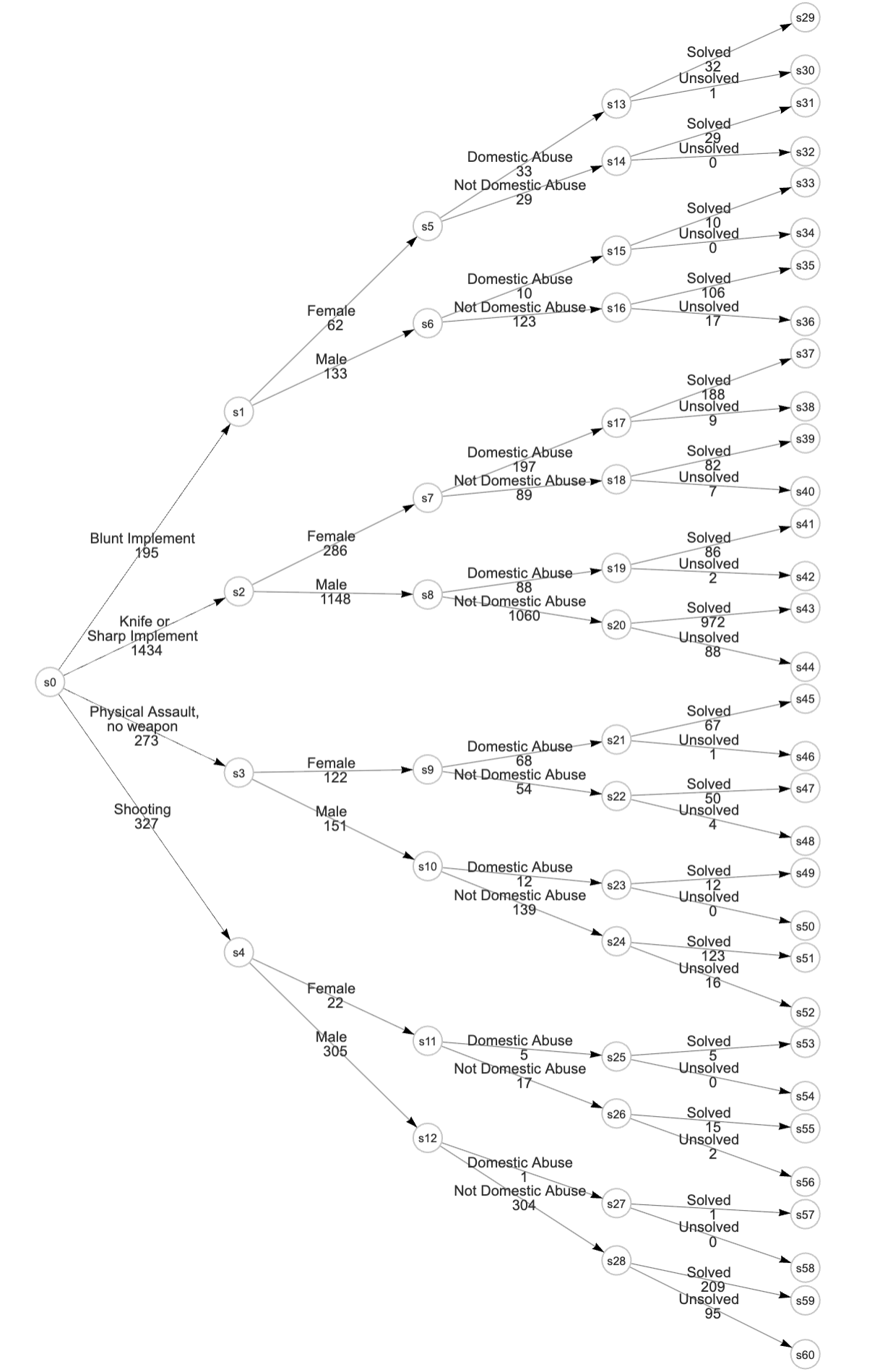}
    \caption{Event Tree for the homicides dataset (\texttt{homicides\_ET} in Section \ref{subsec:formET}). Counts are present on each edge.}
    \label{fig:HomicideET}
\end{figure*}

We can construct the event tree in Figure \ref{fig:HomicideET} from the homicides dataset by using the following code:

\texttt{homicides$\_$ET <- create$\_$event$\_$tree(dataset = homicides, columns = c(\textcolor{blue}{1}:\textcolor{blue}{4}), \\level$\_$separation = \textcolor{blue}{1300}, node$\_$distance = \textcolor{blue}{300}, label$\_$type = \textcolor{ForestGreen}{"both"})}

The user can filter/reorder the variables in their event tree by either manipulating the columns in their dataframe prior to using the function, or by specifying a subset/different ordering in the \texttt{columns} parameter. The \texttt{label$\_$type} parameter specifies how the edge labels will be displayed on the event tree, and can take either: \texttt{"names"}, showing just the variable names, or \texttt{"both"}, which adds the counts of data rows on each edge. The default is \texttt{"both"}.

\newpage 

As mentioned in Section \ref{subsec:createET}, \textbf{stCEG} can support structural asymmetry through the ability to delete nodes in the tree. If this example required an asymmetric event tree, say we only wished to consider whether a case was classed as ``Domestic Abuse'' conditional on the victim being female, then the subset of nodes $\{s_{15}, s_{16}, s_{19}, s_{20}, s_{23}, s_{24}, s_{27}, s_{28}\}$ could be deleted without any loss of information. This would be performed by running:

\texttt{homicides$\_$FemDA <- delete$\_$nodes(homicides$\_$ET, c(\textcolor{ForestGreen}{"s15", "s16", "s19", "s20",\\ "s23", "s24", "s27", "s28"}))}

\begin{figure*}[ht!]
    \centering  
    \includegraphics[width=0.7\textwidth]{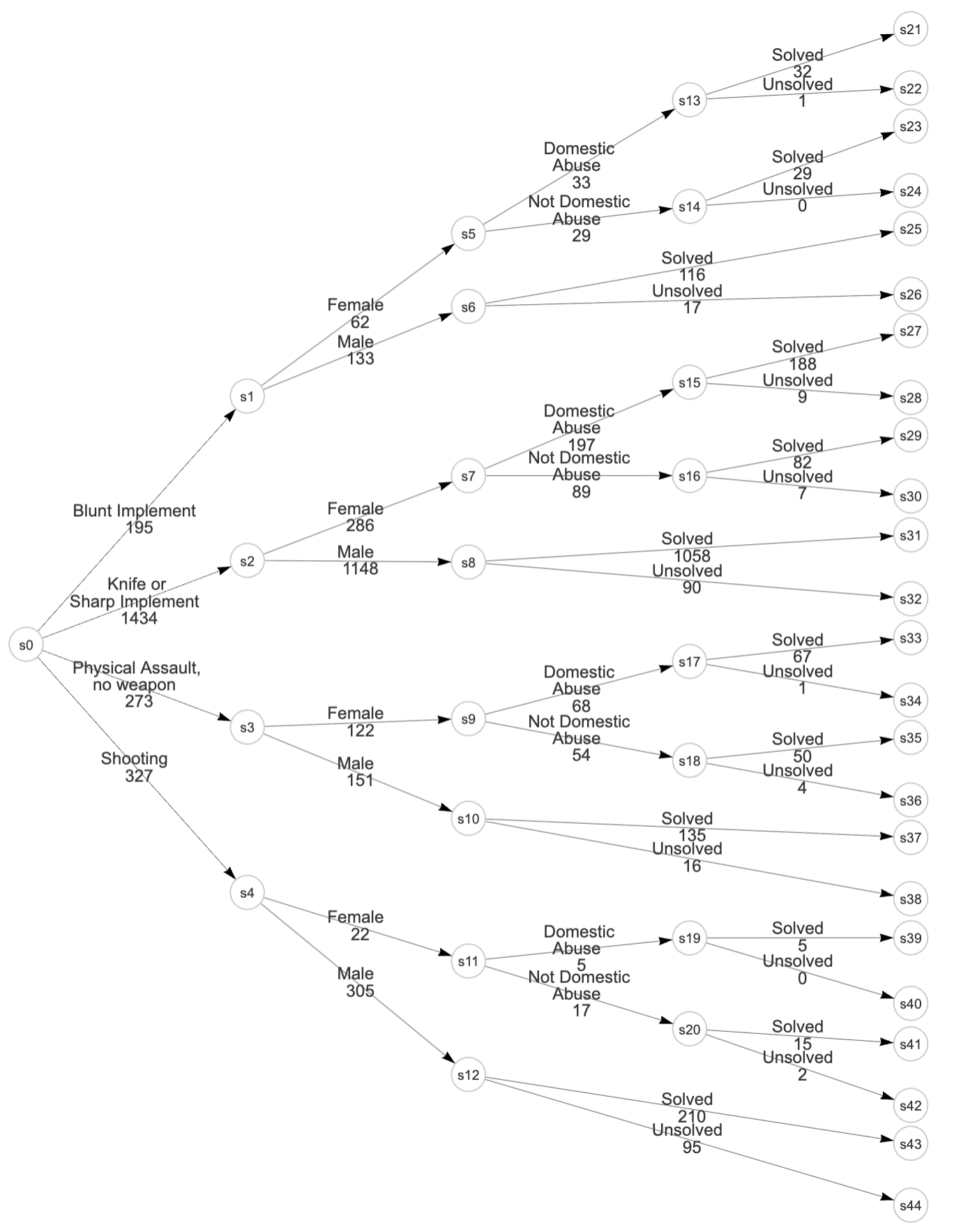}
    \caption{Asymmetric Event Tree for the homicides dataset (\texttt{homicides\_FemDA} in Section \ref{subsec:formET}). Counts are present on each edge.}
    \label{fig:HomicideETAsymmetric}
\end{figure*}

The \texttt{summary} function can be used to get an outline of the structure of the event tree - the number of nodes, edges along with a list of the unique edge labels.

\texttt{summary(homicides\_FemDA)}\\

\texttt{Summary of Nodes}\\
\texttt{================} \\
\texttt{Number of nodes: 45}\\
\texttt{Unique node levels: 5}\\

\texttt{Summary of Edges}\\
\texttt{================} \\
\texttt{Number of edges: 44}\\
\texttt{Unique labels in edges: 10}\\

\texttt{Edge labels}\\
\texttt{===========} \\
\texttt{Blunt Implement}\\
\texttt{Knife or Sharp Implement}\\
\texttt{Physical Assault, no weapon}\\
\texttt{Shooting}\\
\texttt{Female}\\
\texttt{Male}\\
\texttt{Solved}\\
\texttt{Unsolved}\\
\texttt{Domestic Abuse}\\
\texttt{Not Domestic Abuse}\\

This allows users to double-check that the model reflects the structure they intended, and that there aren't any unexpected elements in the structure or event labels.

\subsection{Constructing the Staged Tree: Colouring}
\label{subsec:colouringtree}

Next, we express probability judgements on the event tree. As discussed, the user does not have to colour all vertices using their judgements, the idea being if they are uncertain, the AHC algorithm may be used to complete the colouring. The functions \texttt{update$\_$node$\_$colours} and \texttt{ahc$\_$colouring} assist with this, and can be used in either order. 

The \texttt{update$\_$node$\_$colours} function takes a list of \texttt{node$\_$groups} and a vector of corresponding colours \texttt{colours} as inputs, and can be run as follows:

\texttt{groups <- list(c(\textcolor{ForestGreen}{"s13"}, \textcolor{ForestGreen}{"s21"}), c(\textcolor{ForestGreen}{"s5"}, \textcolor{ForestGreen}{"s9"}), \textcolor{ForestGreen}{"s17"}, c(\textcolor{ForestGreen}{"s25"}), c(\textcolor{ForestGreen}{"s6"},\textcolor{ForestGreen}{"s8"}, \textcolor{ForestGreen}{"s10"}), \textcolor{ForestGreen}{"s12"}, \textcolor{ForestGreen}{"s2"}, \textcolor{ForestGreen}{"s4"})}

\texttt{colour$\_$palette <-c("\colorbox{92DCE5}{\#92DCE5}","\colorbox{C5D86D}{\#C5D86D}","\colorbox{F2DC5D}{\#F2DC5D}","\colorbox{388697}{\#388697}","\colorbox{FE5F55}{\#FE5F55}", "\colorbox{FFAA00}{\#FFAA00}", "\colorbox{A9E5BB}{\#A9E5BB}", "\colorbox{E79C9C}{\#E79C9C}")}

\texttt{homicides$\_$ET$\_$Colour <- update$\_$node$\_$colours(homicides$\_$ET, node$\_$groups = groups, colours = colour$\_$palette)}

\begin{figure*}[ht!]
    \centering  
    \includegraphics[width=0.68\textwidth]{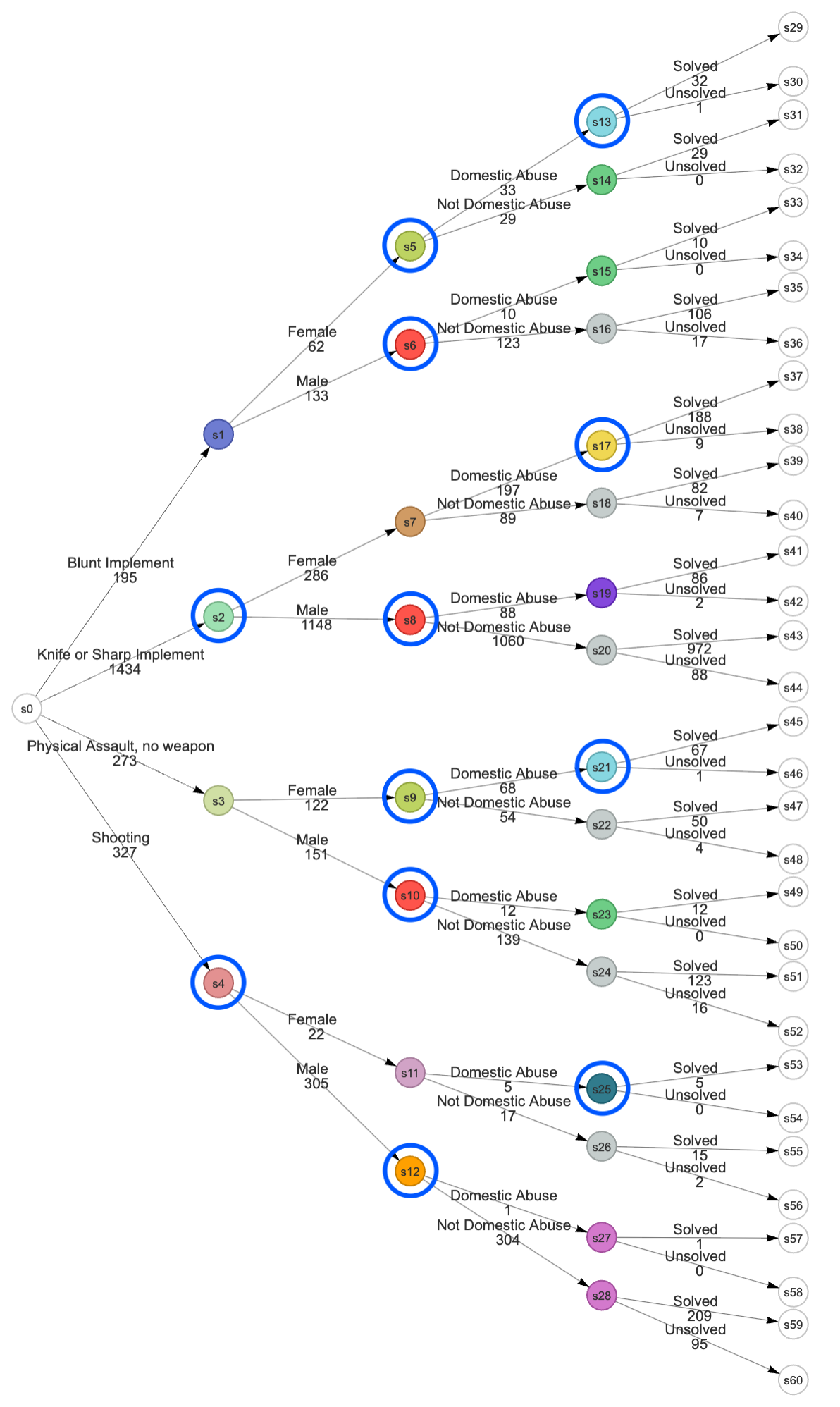}
    \caption{Coloured Event Tree for the homicides dataset. Nodes coloured using user judgements are circled in blue, with the remaining nodes coloured using the AHC algorithm.}
    \label{fig:HomicideAHCET}
\end{figure*}

Figure \ref{fig:HomicideAHCET} shows the nodes coloured using the above code circled in blue. As there are still vertices which aren't coloured, the function \texttt{ahc$\_$colouring} can be used to apply the Agglomerative Hierarchical Clustering algorithm to the remaining nodes. These AHC coloured nodes are uncircled in Figure \ref{fig:HomicideAHCET}:

\texttt{homicides$\_$ET$\_$AHC <- ahc$\_$colouring(homicides$\_$ET$\_$Colour)}

In cases where the user has partially coloured the tree, the summary function can be used to print a description of the current colouring, detailing how many nodes have been coloured and how many need to be coloured before the user is allowed to specify priors:

\texttt{summary(homicides$\_$ET$\_$Colour)}

\vspace{-0.1em}

\texttt{Summary of Staged Tree Object}\\
\texttt{=============================}\\
\texttt{Total nodes: 61}\\
\texttt{Total edges: 60}\\
\texttt{Nodes left to be coloured: 16}

\texttt{Node colour counts:}
\begin{table}[h!]
\begin{tabular}{clclcl}
\textcolor[HTML]{388697}{\rule{1cm}{0.7em}} & \texttt{\#388697} & \texttt{(1 node)} \\
\textcolor[HTML]{92DCE5}{\rule{1cm}{0.7em}} & \texttt{\#92DCE5} & \texttt{(2 nodes)} \\
\textcolor[HTML]{A9E5BB}{\rule{1cm}{0.7em}} & \texttt{\#A9E5BB} & \texttt{(1 node)} \\
\textcolor[HTML]{C5D86D}{\rule{1cm}{0.7em}} & \texttt{\#C5D86D} & \texttt{(2 nodes)} \\
\textcolor[HTML]{E79C9C}{\rule{1cm}{0.7em}} & \texttt{\#E79C9C} & \texttt{(1 node)} \\
\textcolor[HTML]{F2DC5D}{\rule{1cm}{0.7em}} & \texttt{\#F2DC5D} & \texttt{(1 node)} \\
\textcolor[HTML]{FE5F55}{\rule{1cm}{0.7em}} & \texttt{\#FE5F55} & \texttt{(3 nodes)} \\
\textcolor[HTML]{FFAA00}{\rule{1cm}{0.7em}} & \texttt{\#FFAA00} & \texttt{(1 node)} \\
\textcolor[HTML]{FFFFFF}{\rule{1cm}{0.7em}} & \texttt{\#FFFFFF} & \texttt{(49 nodes)} 
\end{tabular}
\vspace{-1em}
\end{table}

\subsection{Constructing the Staged Tree: Prior Specification} \label{subsec:specify_priors}
Through colouring the tree using either of the functions described in Section \ref{subsec:colouringtree} we now have a \texttt{staged$\_$tree} output. Before the conversion of the staged tree into a CEG, we must specify prior judgements for each of our stages. \textbf{stCEG} has 3 types of priors encoded, described individually in Section \ref{subsec:PriorSpec}. The code below allows the user to specify a prior for each stage: 

\texttt{priors <- specify$\_$priors(homicides$\_$ET$\_$AHC, \textcolor{Green}{"Custom"})}

\texttt{Stage Colour Key:}
\vspace{-0.5em}
\begin{table}[htbp]
\begin{tabular}{clclclcl}
\cellcolor[HTML]{FFFFFF} \hspace{1em} & \texttt{\#FFFFFF} &
\cellcolor[HTML]{7987D7} \hspace{1em} & \texttt{\#7987D7} &
\cellcolor[HTML]{A9E5BB} \hspace{1em} & \texttt{\#A9E5BB} &
\cellcolor[HTML]{E79C9C} \hspace{1em} & \texttt{\#E79C9C} \\

\cellcolor[HTML]{D6E3AE} \hspace{1em} & \texttt{\#D6E3AE} &
\cellcolor[HTML]{C5D86D} \hspace{1em} & \texttt{\#C5D86D} &
\cellcolor[HTML]{D6A56F} \hspace{1em} & \texttt{\#D6A56F} &
\cellcolor[HTML]{D7ADCD} \hspace{1em} & \texttt{\#D7ADCD} \\

\cellcolor[HTML]{FE5F55} \hspace{1em} & \texttt{\#FE5F55} &
\cellcolor[HTML]{FFAA00} \hspace{1em} & \texttt{\#FFAA00} &
\cellcolor[HTML]{388697} \hspace{1em} & \texttt{\#388697} &
\cellcolor[HTML]{79D391} \hspace{1em} & \texttt{\#79D391} \\

\cellcolor[HTML]{8F52E0} \hspace{1em} & \texttt{\#8F52E0} &
\cellcolor[HTML]{92DCE5} \hspace{1em} & \texttt{\#92DCE5} &
\cellcolor[HTML]{CCD3D2} \hspace{1em} & \texttt{\#CCD3D2} &
\cellcolor[HTML]{D884D2} \hspace{1em} & \texttt{\#D884D2} \\

\cellcolor[HTML]{F2DC5D} \hspace{1em} & \texttt{\#F2DC5D} & & & & & & \\
\end{tabular}
\vspace{-0.8em}
\end{table}

The user is given a key to match the stage colours to the table if \texttt{print$\_$colours = \textcolor{Periwinkle}{TRUE}}, as R doesn't support printing colour codes in a table. This allows the user to easily see which colour corresponds to each stage, since they can also see the staged tree in the viewer. Once all the priors have been entered, this prints out a table with these filled in (see Table \ref{tab:stprior}). 

\begin{table}[htbp]
\begin{tabular}{llllll}
\texttt{Stage} & \texttt{Colour} & \texttt{Level} & \texttt{Outgoing Edges} & \texttt{Nodes} & \texttt{Prior} \\
\texttt{u1}   & \texttt{\#FFFFFF} & \texttt{1} & \texttt{4} & \texttt{1} & \texttt{""} \\
\texttt{u2}   & \texttt{\#7987d7} & \texttt{2} & \texttt{2} & \texttt{1} & \texttt{""} \\
\texttt{u3}   & \texttt{\#A9E5BB} & \texttt{2} & \texttt{2} & \texttt{1} & \texttt{""} \\
\texttt{u4}   & \texttt{\#E79C9C} & \texttt{2} & \texttt{2} & \texttt{1} & \texttt{""} \\
\texttt{u5}   & \texttt{\#d6e3ae} & \texttt{2} & \texttt{2} & \texttt{1} & \texttt{""} \\
\texttt{u6}   & \texttt{\#C5D86D} & \texttt{3} & \texttt{2} & \texttt{2} & \texttt{""} \\
\texttt{u7}   & \texttt{\#d6a56f} & \texttt{3} & \texttt{2} & \texttt{1} & \texttt{""} \\
\texttt{u8}   & \texttt{\#d7adcd} & \texttt{3} & \texttt{2} & \texttt{1} & \texttt{""} \\
\texttt{u9}   & \texttt{\#fe5f55} & \texttt{3} & \texttt{2} & \texttt{3} & \texttt{""} \\
\texttt{u10}  & \texttt{\#ffaa00} & \texttt{3} & \texttt{2} & \texttt{1} & \texttt{""} \\
\texttt{u11}  & \texttt{\#388697} & \texttt{4} & \texttt{2} & \texttt{1} & \texttt{""} \\
\texttt{u12}  & \texttt{\#79d391} & \texttt{4} & \texttt{2} & \texttt{3} & \texttt{""} \\
\texttt{u13}  & \texttt{\#8f52e0} & \texttt{4} & \texttt{2} & \texttt{1} & \texttt{""} \\
\texttt{u14}  & \texttt{\#92dce5} & \texttt{4} & \texttt{2} & \texttt{2} & \texttt{""} \\
\texttt{u15}  & \texttt{\#ccd3d2} & \texttt{4} & \texttt{2} & \texttt{6} & \texttt{""} \\
\texttt{u16}  & \texttt{\#d884d2} & \texttt{4} & \texttt{2} & \texttt{2} & \texttt{""} \\
\texttt{u17}  & \texttt{\#f2dc5d} & \texttt{4} & \texttt{2} & \texttt{1} & \texttt{""} \\
\end{tabular}

\vspace{1em}
\parbox{0.78\linewidth}{ 
\raggedright
\texttt{Enter new prior for row 1 (4 values, comma-separated):} \\
\vspace{0.1em}
\texttt{\textcolor{blue}{1: 200,1000,400,100}} \\
\vspace{0.1em}
\texttt{Enter new prior for row 2 (2 values, comma-separated):} \\
\texttt{\textcolor{blue}{1: 25,75}} \\
\vspace{0.1em}
\texttt{Enter new prior for row 3 (2 values, comma-separated):} \\
\texttt{\textcolor{blue}{1: 300,900}} \\
\vspace{0.1em}
\texttt{Enter new prior for row 4 (2 values, comma-separated):} \\
\texttt{\textcolor{blue}{1: 50,50}} \\
\vspace{0.1em}
}
\vspace{-1em}
\end{table}

Alternatively we can start with an uninformative prior for all stages and edit specific priors if desired via the console, by setting \texttt{ask$\_$edit = \textcolor{Periwinkle}{TRUE}/\textcolor{Periwinkle}{FALSE}} (this editing process is shown in Section \ref{subsec:PriorSpec}).

Given we have specified a prior for each of our stages, this dataframe (\texttt{priors} in the example) then becomes an input together with a \texttt{staged$\_$tree} object in the \texttt{staged$\_$tree$\_$prior} function. This function links the priors together with the tree structure to form a staged tree. The options for \texttt{label$\_$type} are \texttt{priors}, \texttt{prior$\_$means} or \texttt{none}, with displayed value replacing the data values as labels as in Figure \ref{fig:HomicideAHCET}. 

\texttt{homicides$\_$ST <- staged$\_$tree$\_$prior(homicides$\_$ET$\_$AHC, priors, label$\_$type = \textcolor{ForestGreen}{"priors"})}

\begin{table}[h!]
\centering
\resizebox{0.95\textwidth}{!}{
\begin{tabular}{p{1cm} p{1.5cm} p{1cm} p{2.7cm} p{1cm} p{3cm} p{3cm}}
\texttt{Stage} & \texttt{Colour} & \texttt{Level} & \texttt{Outgoing Edges} & \texttt{Nodes} & \texttt{Prior} & \texttt{Prior Mean} \\ 
\texttt{u1} & \texttt{\#FFFFFF} & \texttt{1} & \texttt{4} & \texttt{1} & \texttt{200,1000,400,100} & \texttt{0.12,0.59,0.24,0.06} \\
\texttt{u2} & \texttt{\#7987D7} & \texttt{2} & \texttt{2} & \texttt{1} & \texttt{25,75} & \texttt{0.25,0.75} \\
\texttt{u3} & \texttt{\#A9E5BB} & \texttt{2} & \texttt{2} & \texttt{1} & \texttt{300,900} & \texttt{0.25,0.75} \\
\texttt{u4} & \texttt{\#D6E3AE} & \texttt{2} & \texttt{2} & \texttt{1} & \texttt{50,50} & \texttt{0.5,0.5} \\
\texttt{u5} & \texttt{\#E79C9C} & \texttt{2} & \texttt{2} & \texttt{1} & \texttt{10,140} & \texttt{0.07,0.93} \\
\texttt{u6} & \texttt{\#C5D86D} & \texttt{3} & \texttt{2} & \texttt{2} & \texttt{20,10} & \texttt{0.67,0.33} \\
\texttt{u7} & \texttt{\#D6A56F} & \texttt{3} & \texttt{2} & \texttt{1} & \texttt{60,40} & \texttt{0.6,0.4} \\
\texttt{u8} & \texttt{\#D7ADCD} & \texttt{3} & \texttt{2} & \texttt{1} & \texttt{3,2} & \texttt{0.6,0.4} \\
\end{tabular}
}
\end{table}

\begin{table}[h!]
\centering
\resizebox{0.95\textwidth}{!}{
\begin{tabular}{p{1cm} p{1.5cm} p{1cm} p{2.7cm} p{1cm} p{3cm} p{3cm}}
\texttt{Stage} & \texttt{Colour} & \texttt{Level} & \texttt{Outgoing Edges} & \texttt{Nodes} & \texttt{Prior} & \texttt{Prior Mean} \\ 
\texttt{u9} & \texttt{\#FE5F55} & \texttt{3} & \texttt{2} & \texttt{3} & \texttt{50,950} & \texttt{0.05,0.95} \\
\texttt{u10} & \texttt{\#FFAA00} & \texttt{3} & \texttt{2} & \texttt{1} & \texttt{1,99} & \texttt{0.01,0.99} \\
\texttt{u11} & \texttt{\#388697} & \texttt{4} & \texttt{2} & \texttt{1} & \texttt{10,0} & \texttt{1,0} \\
\texttt{u12} & \texttt{\#79D391} & \texttt{4} & \texttt{2} & \texttt{3} & \texttt{5,1} & \texttt{0.83,0.17} \\
\texttt{u13} & \texttt{\#8F52E0} & \texttt{4} & \texttt{2} & \texttt{1} & \texttt{90,5} & \texttt{0.95,0.05} \\
\texttt{u14} & \texttt{\#92DCE5} & \texttt{4} & \texttt{2} & \texttt{2} & \texttt{50,3} & \texttt{0.94,0.06} \\
\texttt{u15} & \texttt{\#CCD3D2} & \texttt{4} & \texttt{2} & \texttt{6} & \texttt{30,8} & \texttt{0.79,0.21} \\
\texttt{u16} & \texttt{\#D884D2} & \texttt{4} & \texttt{2} & \texttt{2} & \texttt{70,65} & \texttt{0.52,0.48} \\
\texttt{u17} & \texttt{\#F2DC5D} & \texttt{4} & \texttt{2} & \texttt{1} & \texttt{12,4} & \texttt{0.75,0.25} \\
\end{tabular}
}
\vspace{0.5em}
\caption{Custom user-specified priors for the staged tree shown in Figure \ref{fig:HomicideAHCET}.}
\label{tab:stprior}
\vspace{-2em}
\end{table}

\subsection{Creating the Chain Event Graph}

When the prior judgements have been linked to the staged tree using the \texttt{staged$\_$tree$\_$prior} function, this output can now be passed to the \texttt{create$\_$ceg} function. The data values along with the user-specified priors are used to calculate posterior distributions for each of the stages. These updates can be shown in the form of a table by setting \texttt{view$\_$table = TRUE} in \texttt{create$\_$ceg}. When plotting the CEG, the category labels (e.g. ``Female'') are automatically displayed. The user has the ability to create a prior CEG by setting \texttt{label = ""} to display either prior parameter values (\texttt{prior}) or means (\texttt{prior$\_$mean}) for each stage, or a posterior CEG with posterior parameter values (\texttt{posterior}) or means (\texttt{posterior$\_$mean}). There is also the option to completely omit the numeric labels and only show the structure, using \texttt{label = "none"}.

\texttt{homicides$\_$CEG <- create$\_$ceg(homicides$\_$ST, view$\_$table = TRUE, label = \\\textcolor{ForestGreen}{"posterior$\_$mean"}, level$\_$separation = \textcolor{blue}{1500})}
\vspace{-0.2em}
\begin{figure*}[h!]
    \centering  
    \includegraphics[width=0.74\textwidth]{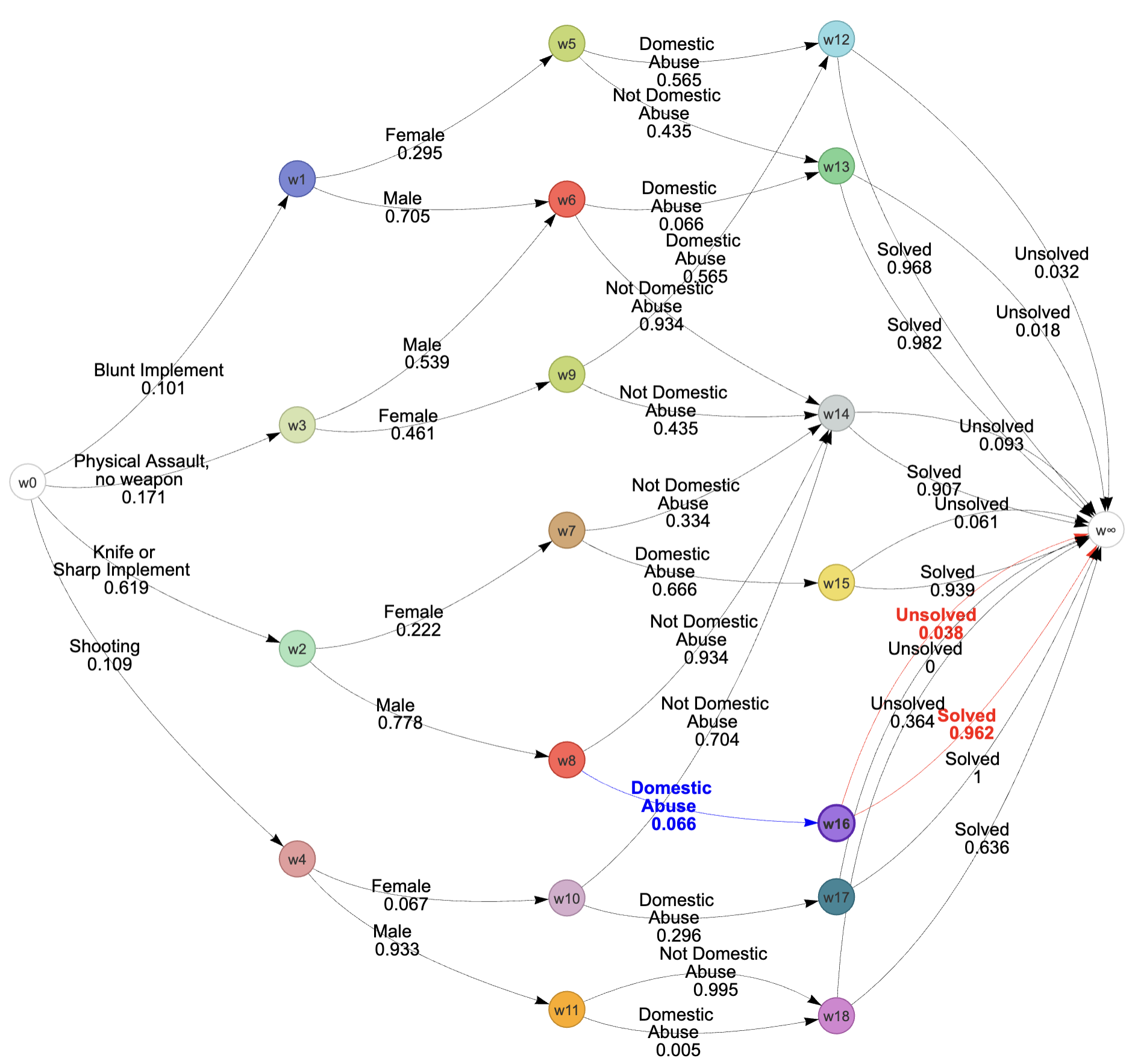}
    \caption{Chain Event Graph for the homicides dataset, with posterior means as edge labels. Note that as $w_{16}$ has been selected, edges entering it are highlighted blue, and those leaving it are highlighted red.}
    \label{fig:HomicideCEG}
\end{figure*}

The CEG plot (Figure \ref{fig:HomicideCEG}) is interactive, so nodes are able to be reordered by dragging them in the viewer. When a node is selected, its incoming edges are highlighted blue, and its outgoing edges red. This makes it simple to track paths through the tree.

\begin{figure*}[ht!]
    \centering  
    \includegraphics[width=0.9\textwidth]{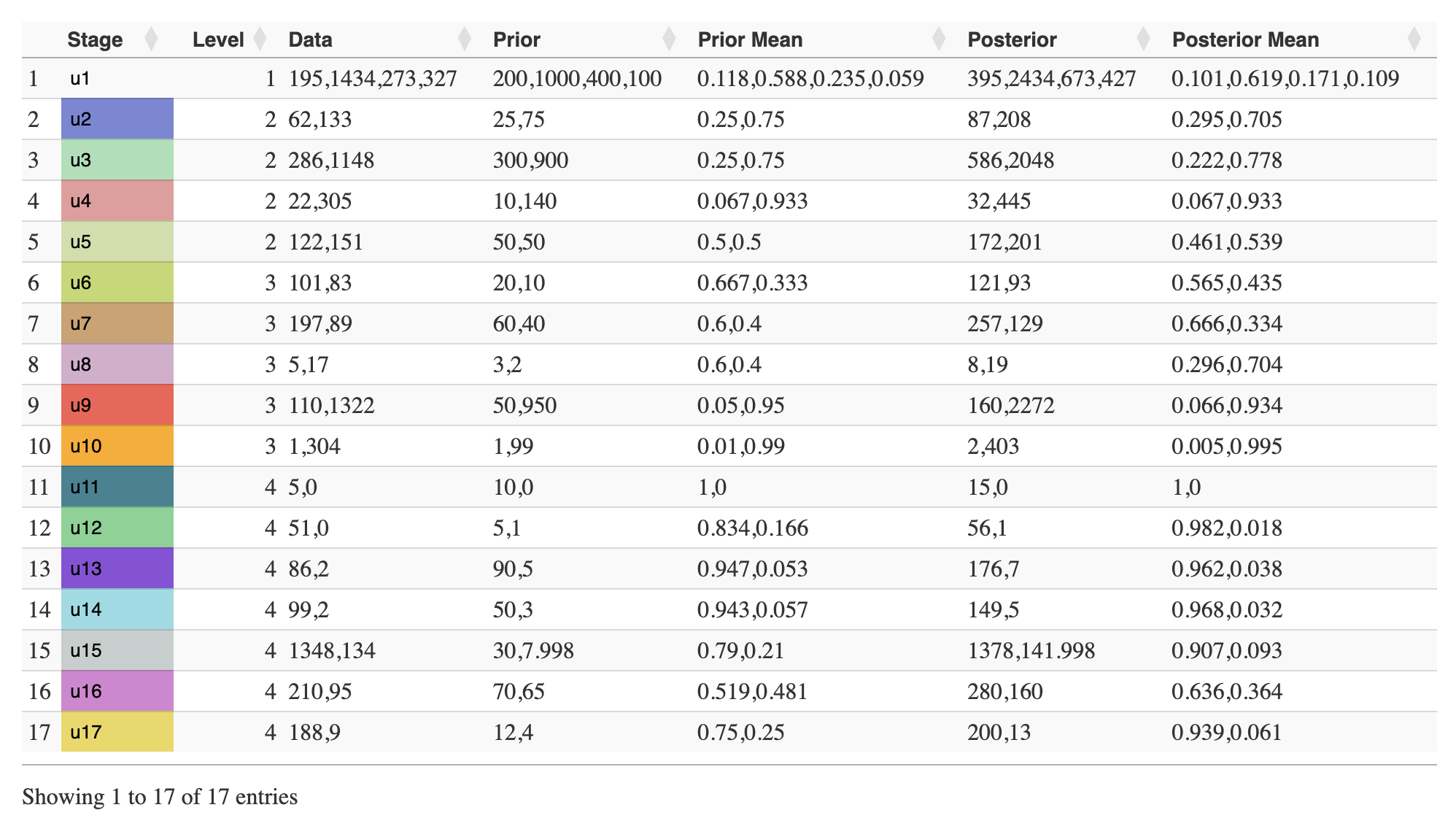}
    \caption{Prior to posterior update table for the CEG shown in Figure \ref{fig:HomicideCEG}, as seen in the RStudio viewer.}
    \label{tab:stcegupdate}
\end{figure*}

As well as being able to view how the distributions for each stage have updated, we can also use the \texttt{summary} function, to calculate the log-marginal likelihood of the model, which is useful in model comparison:

\texttt{summary(homicides$\_$CEG)}
\begin{table}[h!]
\texttt{Chain Event Graph Summary}\\
\texttt{----------------------------------------------------}\\
\texttt{Total Log Marginal Likelihood:  -4938.692}\\
\begin{tabular}{l r r l}
\texttt{Stage} & \texttt{LogScore} & \texttt{ESS} & \\ 
\texttt{u1} & \texttt{-2388.659} & \texttt{3929} & \\
\texttt{u2} & \texttt{-123.277} & \texttt{295} & \\
\texttt{u3} & \texttt{-721.632} & \texttt{2634} & \\
\texttt{u4} & \texttt{-81.206} & \texttt{477} & \\
\texttt{u5} & \texttt{-188.758} & \texttt{373} & \\
\texttt{u6} & \texttt{-128.442} & \texttt{214} & \\
\texttt{u7} & \texttt{-179.263} & \texttt{386} & \\
\texttt{u8} & \texttt{-13.856} & \texttt{27} & \texttt{**} \\
\texttt{u9} & \texttt{-392.084} & \texttt{2432} &\\
\texttt{u10} & \texttt{-7.405} & \texttt{405} & \\
\texttt{u11} & \texttt{-0.693} & \texttt{15} & \texttt{**} \\
\texttt{u12} & \texttt{-2.416} & \texttt{57} & \texttt{**} \\
\texttt{u13} & \texttt{-10.302} & \texttt{183} & \\
\texttt{u14} & \texttt{-10.806} & \texttt{154} & \\
\texttt{u15} & \texttt{-453.740} & \texttt{1520} & \\
\texttt{u16} & \texttt{-195.484} & \texttt{440} & \\
\texttt{u17} & \texttt{-40.668} & \texttt{213} & \\
\end{tabular}
\vspace{-1em}
\end{table}

\texttt{Note: ESS (Effective Sample Size) reflects the total information (prior + data) available for each stage.}\\
\texttt{Stages with ESS < 100 are flagged with ''**`` as potentially low-information stages.}\\
\texttt{Increasing the strength of the prior would help this.}\\

Given we have two competing models using the same data (such as if experts gave differing opinions on an event process), we can use the \texttt{compare$\_$ceg$\_$models} function to quantify using the log-likelihood ratio the support for one model over another. In this example, \texttt{homicides$\_$CEG$\_$AHC} is a completely uninformative model, using AHC for colouring, and a Uniform prior for each stage. Code for this can be found in Appendix \ref{app:comparisoncode}.

\vspace{0.5em}
\texttt{homicides$\_$CEG$\_$summary <- summary(homicides$\_$CEG)}\\
\texttt{homicides$\_$CEG$\_$AHC$\_$summary <- summary(homicides$\_$CEG$\_$AHC)}\\
\texttt{compare$\_$ceg$\_$models(homicides$\_$CEG$\_$summary,homicides$\_$CEG$\_$AHC$\_$summary)}
\vspace{0.5em}

\texttt{Log marginal of model 1:  -4938.692}\\
\texttt{Log marginal of model 2:  -4890.359}\\
\texttt{Log Bayes factor of Model 1 vs Model 2:  -48.333}\\
\texttt{Preferred Model: Model 2}
\vspace{0.5em}

This function uses the same scoring method employed by the Agglomerative Hierarchical Clustering (AHC) algorithm, the details of which can be found in Appendix \ref{app:dirichletcalculation}. The log Bayes factor is used to compare models: a positive value supports Model 1, while a negative value favours Model 2.

It is important to note, however, that this score reflects how well the model’s prior distributions align with the observed data. As such, in scenarios where data is sparse or certain groups are under-represented, or if the user suspects the data may not accurately reflect the underlying system, this scoring metric should be interpreted with caution.

The log marginal likelihood integrates over the uncertainty in the parameters rather than conditioning on point estimates, making it sensitive both to the prior and to the amount of data supporting each stage. In stages with limited observations, the score may reflect the prior more than the data, which can result in overly simplistic models being favoured if their priors happen to align well with sparse observations.
\vspace{-1.2em}
\subsection{Exploring Space with the CEG}
\label{subsec:SpaceCEG}
In the case where we have a CEG of a system that exhibits differences across geographical areas, we may wish to model this variation. As an example by considering homicides as variable over London Boroughs we extend the model in Figure \ref{fig:HomicideCEG}. One way to do this is by adding the spatial variable to the initial tree. However, as there are 32 boroughs, this hugely increases the size of the event tree (from 60 vertices to 1952 vertices), and hence the CEG. To address this, the function \texttt{generate$\_$ceg$\_$map} takes a CEG and corresponding shapefile, using these to generate a map to allow us to leverage the results of the CEG by viewing smaller sections broken down by area. We fit a new CEG with the addition of the Borough column in our dataset using the process described above. The code for this can be found in Appendix \ref{app:spatialcode}. This is named \texttt{homicides$\_$CEG$\_$spatial}. We also have a shapefile \texttt{london$\_$boroughs}, and thus can run the following code:

\texttt{generate$\_$CEG$\_$map(london$\_$boroughs, homicides$\_$CEG$\_$spatial, colour$\_$by = \textcolor{ForestGreen}{"Solved"})
}
\vspace{-0.5em}

\begin{figure}[h!]
\vspace{-1em}
    \centering
    \subfloat[Map generated for CEG model including Borough column.]{%
        \includegraphics[width=0.45\textwidth]{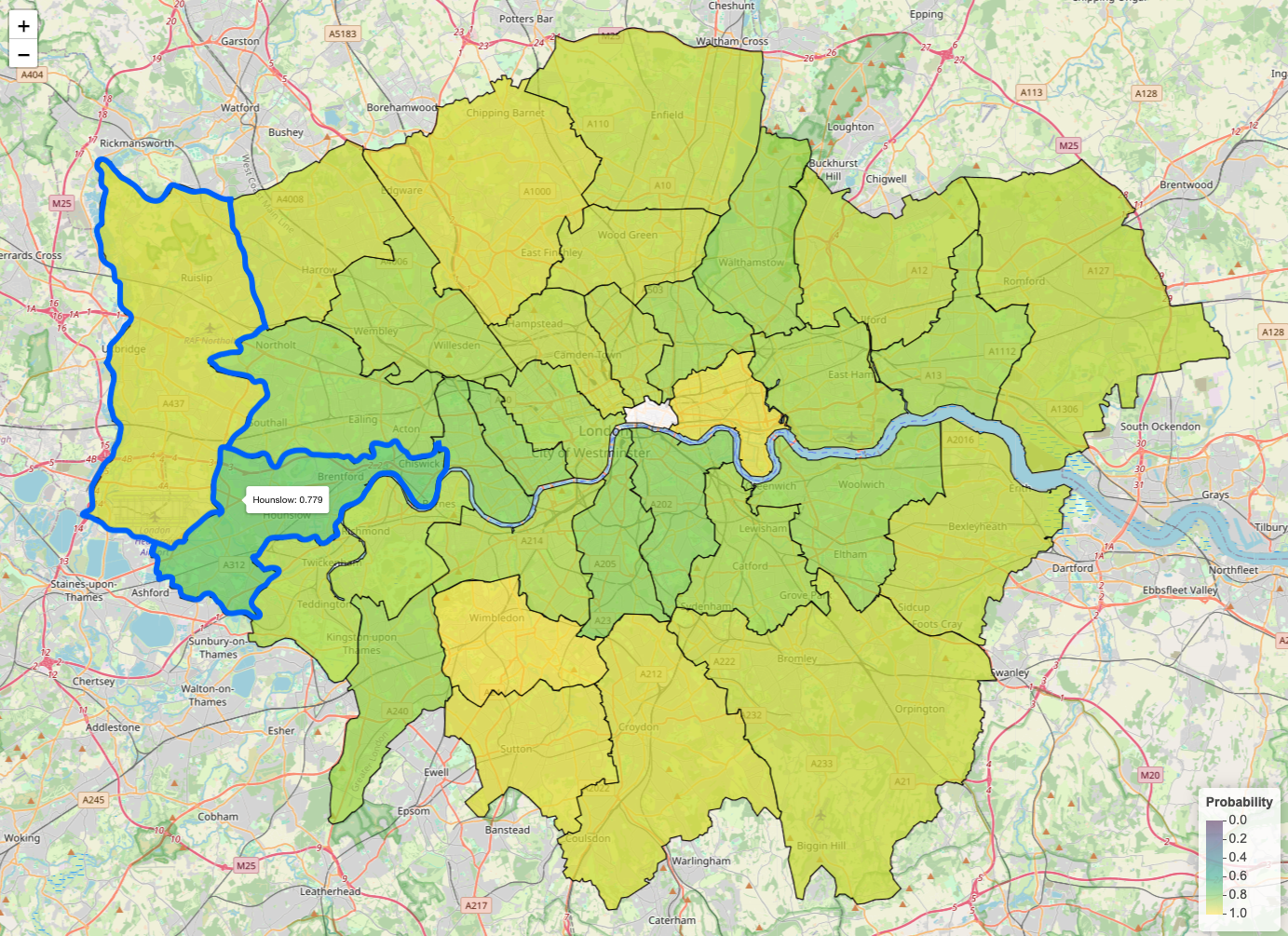}
        \label{fig:cegspacemap}
    }
    \hfill
    \subfloat[Map generated for CEG model including Borough column, conditioned on ``Knife or Sharp Implement'' and ``Female''.]{
    {\includegraphics[width=0.45\textwidth]{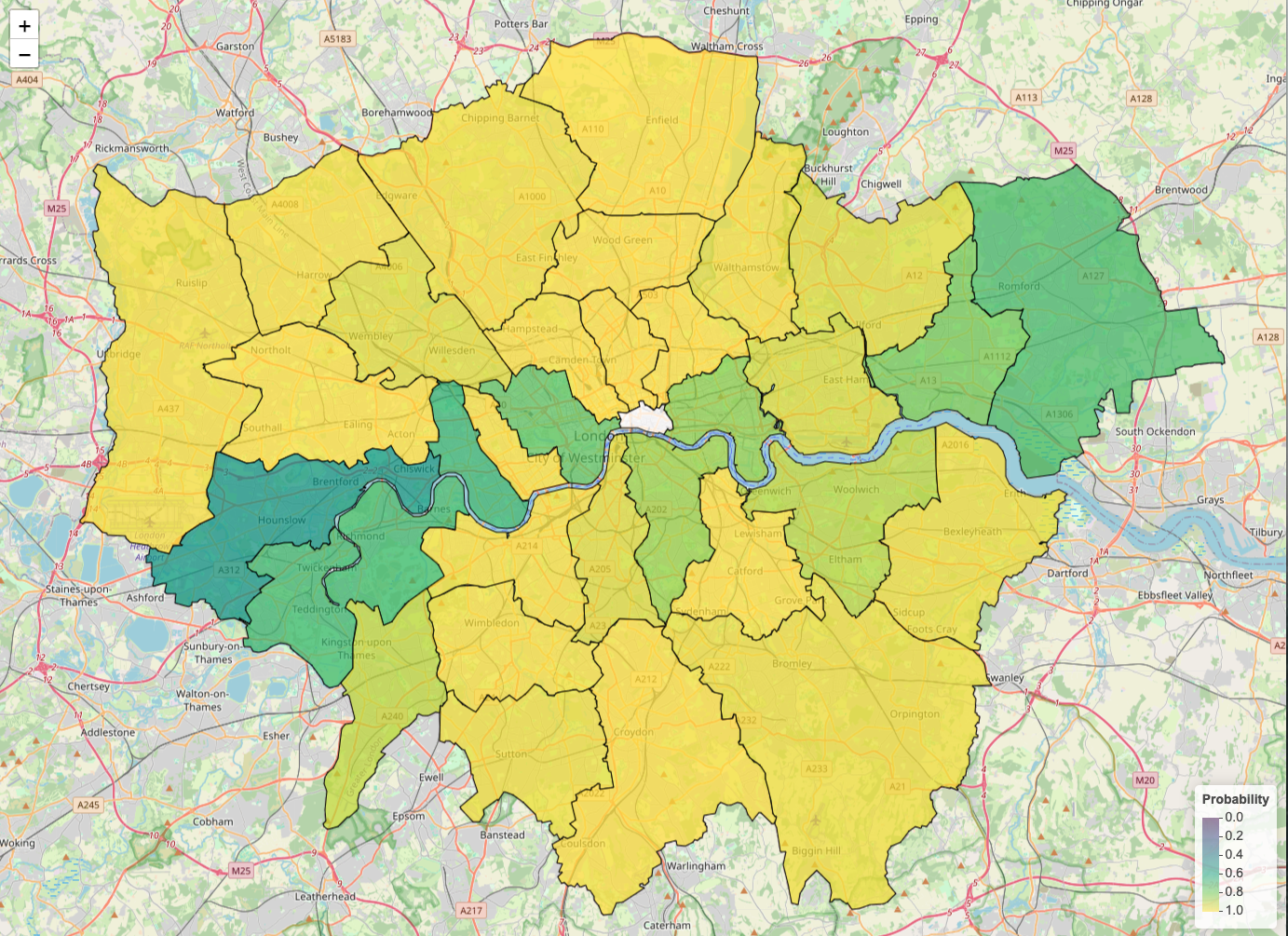}}
        \label{fig:cegspacemapconditionals}
    }
    \caption{Map generated for \texttt{homicides\_CEG\_spatial} model. Area probabilities are shown when polygons are hovered over. Note in Figure \ref{fig:cegspacemap} Hillingdon and Hounslow are outlined in blue.}
    \label{fig:mapandconditionals}
\end{figure}

This code produces the map found in Figure \ref{fig:cegspacemap}. By hovering over each polygon, the user can see the averaged probability from the CEG for that area. The user can also create maps conditioned on specific variable categories by specifying these with the \texttt{conditionals} option in \texttt{generate$\_$CEG$\_$map}.

\texttt{generate$\_$CEG$\_$map(london$\_$boroughs, homicides$\_$CEG$\_$spatial, colour$\_$by = \textcolor{ForestGreen}{"Solved"}, color$\_$palette = \textcolor{ForestGreen}{"viridis"}, conditionals = c(\textcolor{ForestGreen}{"Knife or Sharp Implement"}, \textcolor{ForestGreen}{"Female"}))}

If the user wants to explore the branches of the CEG associated with a specific area to compare the model structure between several areas instead of just the probabilities, then these can be shown using the \texttt{create$\_$reduced$\_$CEG} function. The model results in Hillingdon and Hounslow having very different probabilities of a case being solved despite being adjacent to each other (this can be seen as there is a marked colour difference between these boroughs, outlined in blue in Figure \ref{fig:cegspacemap}). These reduced CEGs could be compared to see which combinations of factors were causing the difference in solved probability:

\texttt{create$\_$reduced$\_$CEG(homicides$\_$CEG$\_$spatial, \textcolor{ForestGreen}{"Hillingdon"}) \\
create$\_$reduced$\_$CEG(homicides$\_$CEG$\_$spatial, \textcolor{ForestGreen}{"Hounslow"})}

\begin{figure}[h!]
    \centering
    \subfloat[Reduced CEG from Hillingdon]{%
        \includegraphics[width=0.485\textwidth]{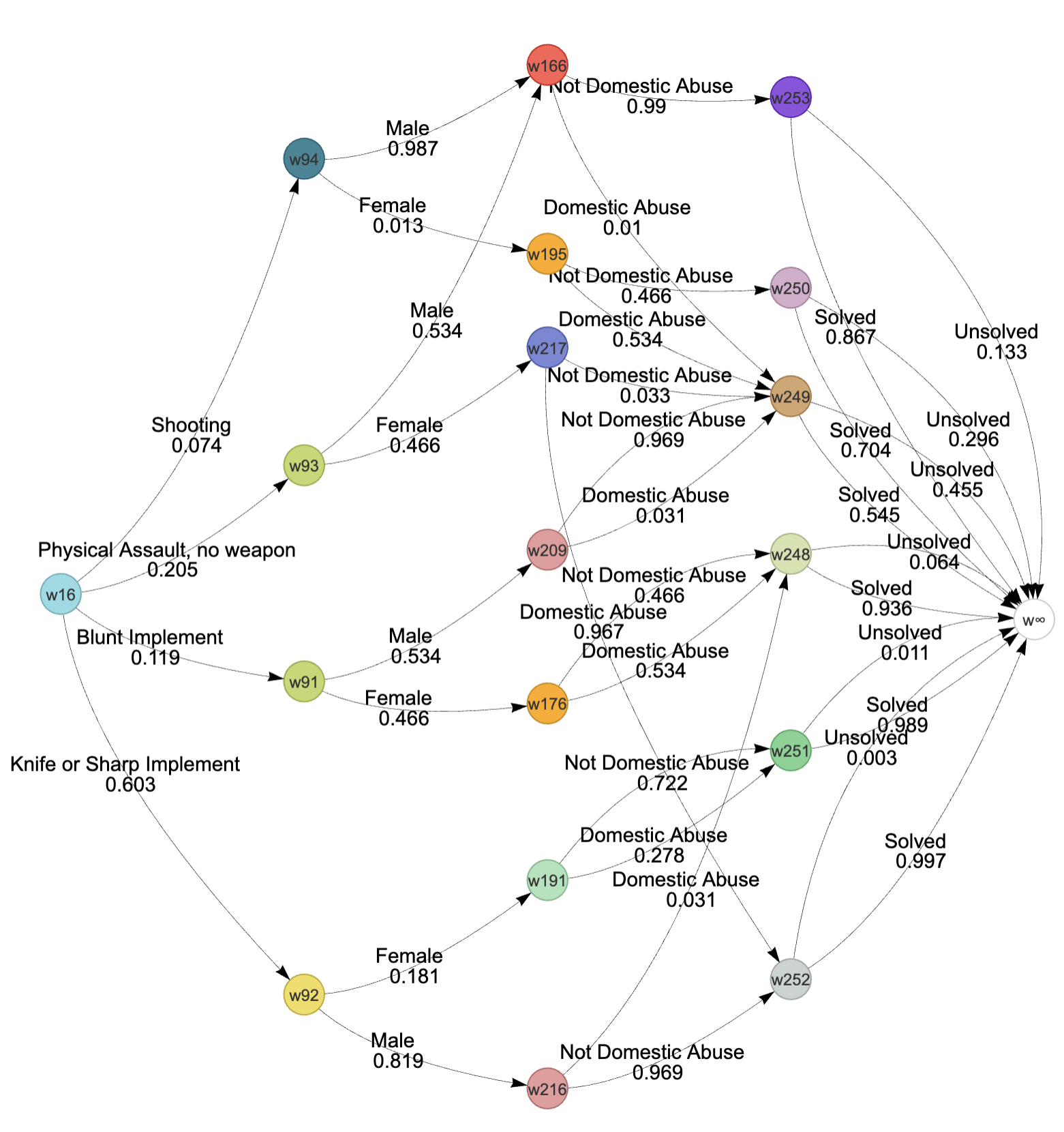}
        \label{fig:hillingdonCEG}
    }
    \hfill
    \subfloat[Reduced CEG from Hounslow]{
    {\includegraphics[width=0.485\textwidth]{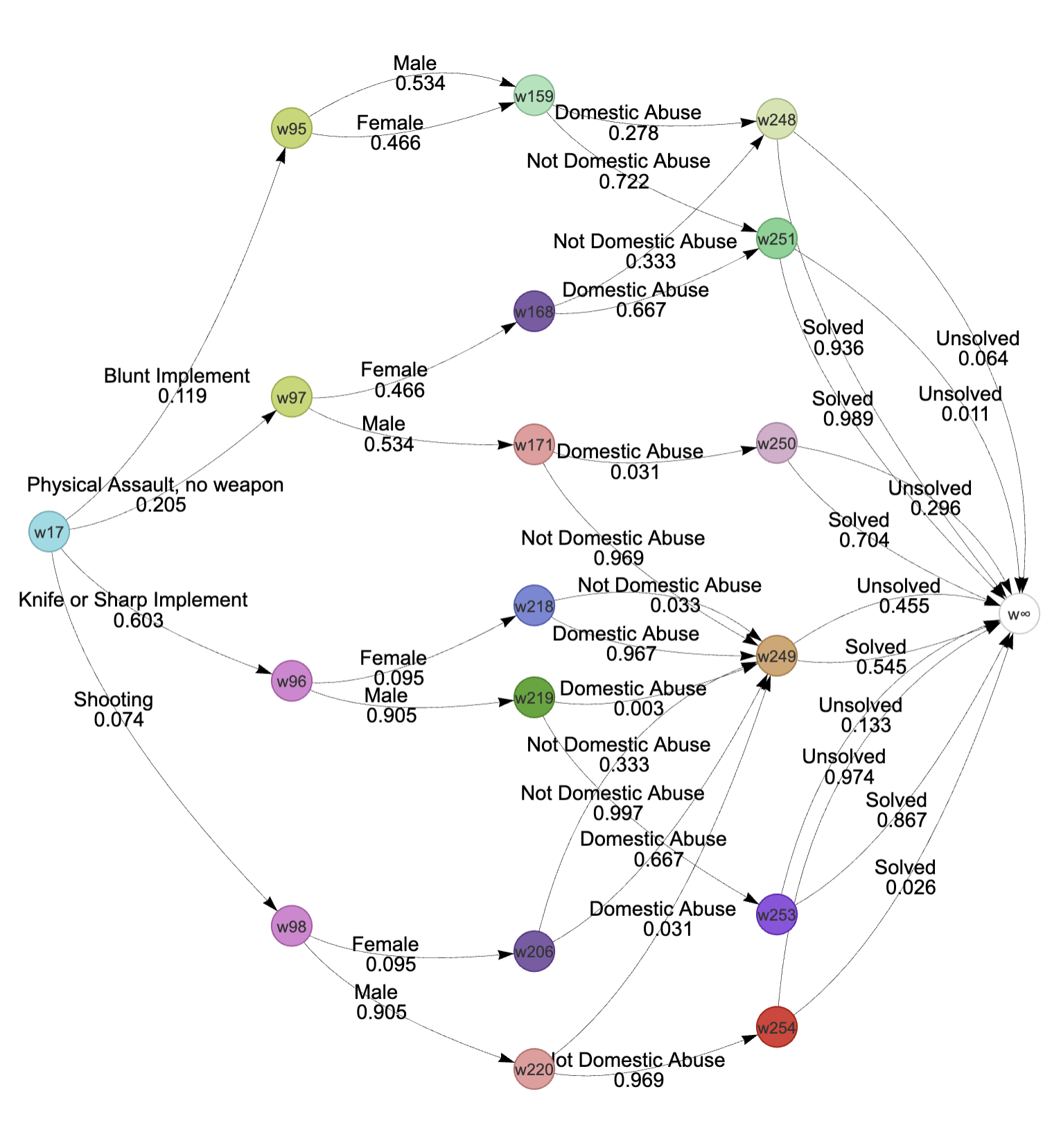}}
        \label{fig:hounslowCEG}
    }
    \vspace{2em}
    \caption{Reduced CEGs for Hillingdon and Hounslow. Note that stage colours match between these reduced CEGs, but not to Figure \ref{fig:HomicideCEG}, as a new larger model has been fit. The code to fit this larger model can be found in Appendix \ref{app:spatialcode}.}
    \label{fig:hillingdonhounslowcomparison}
\end{figure}

\section{The stCEG User Interface}
Chain Event Graphs aim to encode expert judgement without the need for users to have formal mathematical training to understand their output. This motivates the need for a web-based user interface to ensure that a lack of coding experience is not a barrier to entry for these models. The \textbf{stCEG} UI is developed using \textbf{shiny}, and can be launched directly from the R console by calling the function \texttt{run$\_$stceg}. Alternatively, an online version of the \textbf{stCEG} web application can be found at \url{https://holliecalley.shinyapps.io/stceg/}.

The app is organised into 3 tabs, \texttt{Upload Data} , \texttt{Select Data}, and \texttt{Plots}, allowing the user to navigate through the application. The details of each tab are described below.
\subsection{Upload Data}

\begin{figure*}[h!]
    \centering  
    \makebox[0.9\textwidth]{\includegraphics[width=0.95\textwidth]{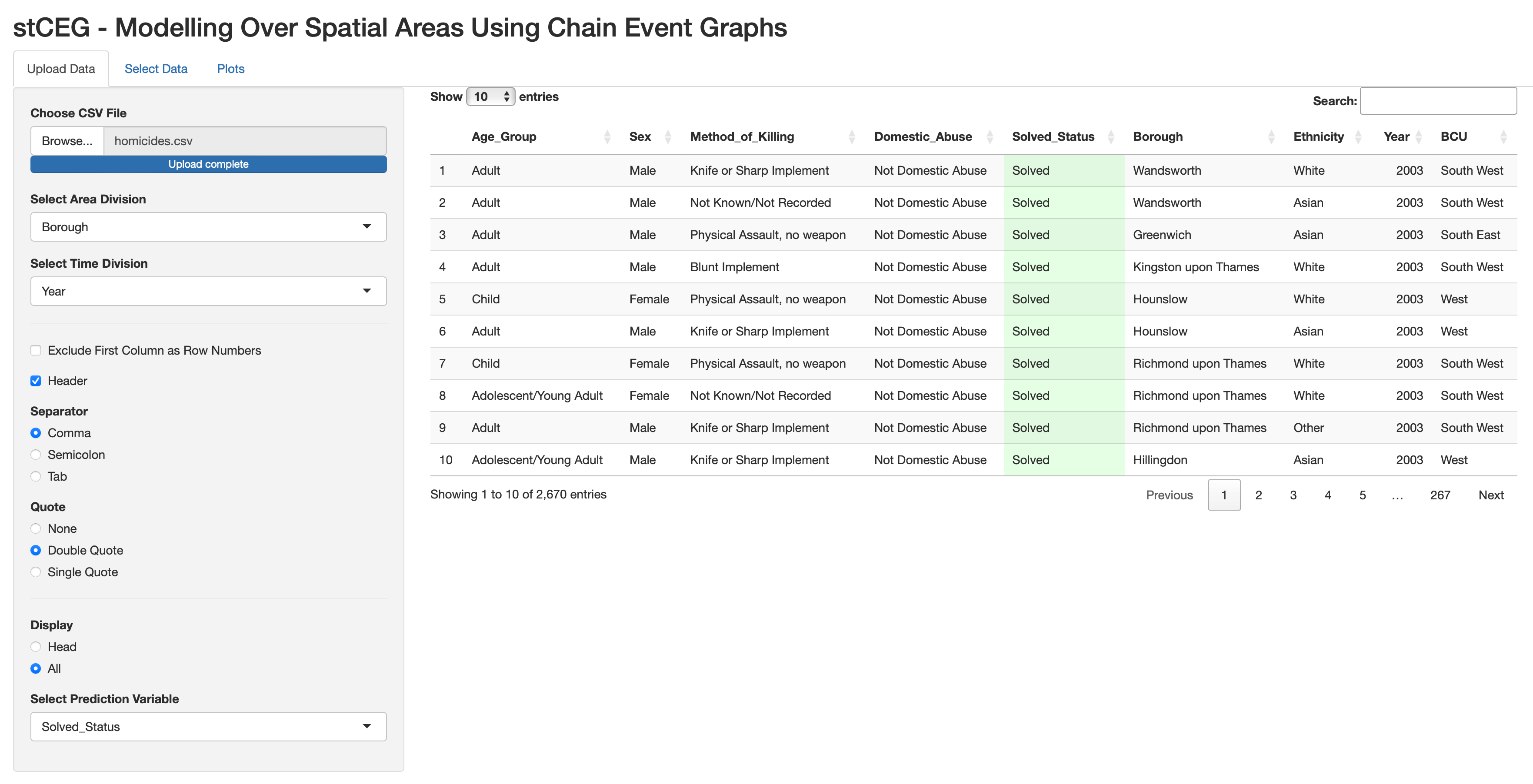}}
    \caption{The \texttt{Upload Data} panel in the \textbf{stCEG} web application.}
    \label{fig:uploaddata}
\end{figure*}

\textbf{Step 1}: In the initial tab (Figure \ref{fig:uploaddata}), the user uploads a dataset in \texttt{csv} form, and then is asked to select if there are any area or time divisions present. Standard options such as quote, separator, and header are available to ensure the data is read correctly. A prediction variable must be specified, which is automatically used as the final variable when the event tree is created. The user can then view the dataset, with the prediction variable column highlighted in green. 
\vspace{-0.5em}
\vspace{-1em}
\subsection{Select Data}

\begin{figure*}[h!]
    \centering
    \makebox[0.9\textwidth]{ 
    \includegraphics[width=0.95\textwidth]{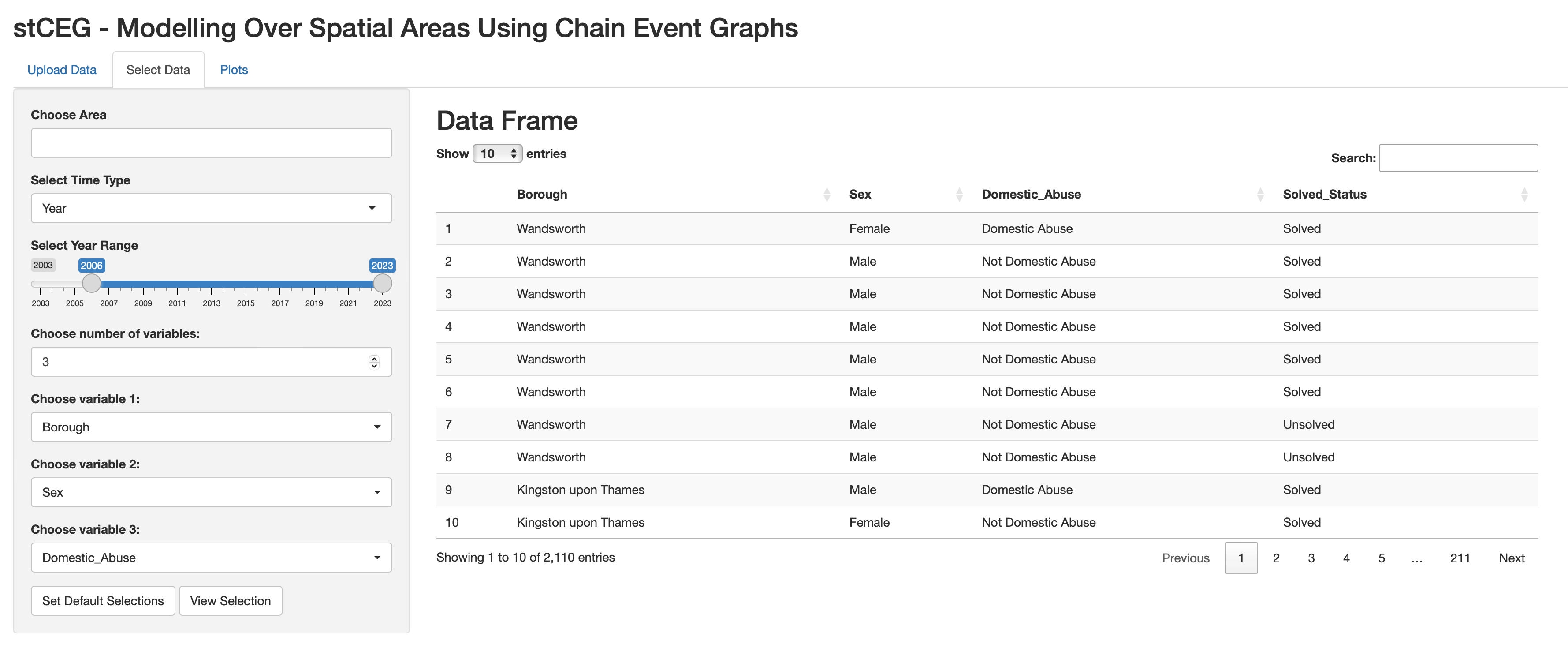}}
    \caption{The \texttt{Select Data} panel in the \textbf{stCEG} web application.}
    \label{fig:selectdata}
\end{figure*}

\textbf{Step 2}: This tab (shown in Figure \ref{fig:selectdata}) is used to filter the data, informing the model structure. If any time or area columns were identified in the previous tab, options will appear to allow the user to use these to filter the data:
\vspace{-0.5em}
\begin{itemize}
    \item If the user has selected an area column, then an option will appear in order to select a subset of these areas. If this is left blank, all areas remain post-filtering.
    \vspace{-0.5em}
    \item Where a time column has been identified, the user must specify its granularity (e.g. year, year-month, date) and how it is formatted. This produces a slider for the user to limit the data to a specific range. 
\end{itemize}
\vspace{-0.5em}

\textbf{Step 3}: 
The user is required to choose the number of variables they wish to include in the model, noting that the prediction variable selected in the previous tab will automatically be the final variable in the tree. Given the number of variables the user chose, each variable is rendered as a drop-down box (shown on the left of Figure \ref{fig:selectdata}). The order these variables are presented mirrors how they will appear in the event tree, so the user should bear that in mind when selecting. Given the number of variables has been chosen, the \texttt{Set Default Selections} button (bottom left) selects this number of variables in the order they appear in the data. If an area column was specified, this option also selects two random areas. Finally, clicking on the \texttt{View Selection} button (right of \texttt{Set Default Selections}) displays the filtered dataframe.

\subsection{Plots}
\label{subsec:plots}
\textbf{Step 4}: 
The user can create the event tree from the specified variable order by clicking on the \texttt{View Event Tree} button. By default, all combinations of characteristics are given as paths in the tree. If the user believes the tree requires an asymmetric structure, they can select all nodes they want to delete and do so using the \texttt{Delete Selected Node} button. 

\textbf{stCEG} provides the user with two methods for colouring the event tree, given that an area variable is present in the filtered data: \texttt{Event Tree}, and \texttt{Map}. \texttt{Event Tree} simply allows the user to zoom in and interact with the tree by clicking nodes to colour them. The \texttt{Choose Colour} option enables the user to choose a custom colour palette, and once a colour is chosen, it can be applied to the selected nodes by clicking \texttt{Update Colour}. Much like the console functions \texttt{update$\_$node$\_$colours} and \texttt{ahc$\_$colouring}, the user can either fully colour every node themselves, use the AHC algorithm over the entire tree (using the \texttt{Colour using AHC}) button), or a combination of both methods in either order. All situations must be coloured before the user is allowed to click the \texttt{Finished Colouring} button.

\begin{figure*}[ht!]
    \centering
    \makebox[\textwidth]{ 
    \includegraphics[width=0.9\textwidth]{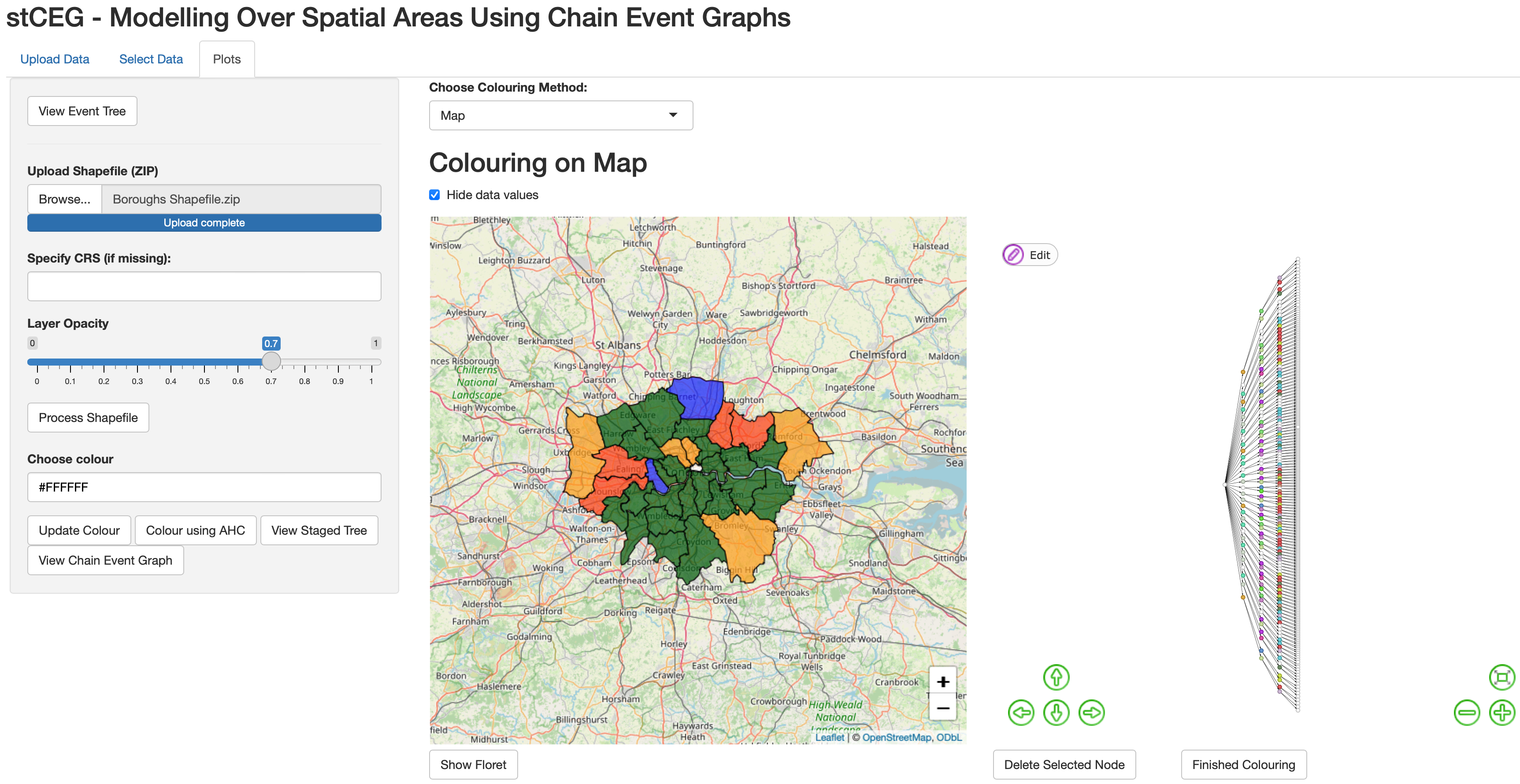}}
    \caption{The \texttt{Plots} panel in the \textbf{stCEG} web application, showing how the user can colour an event tree.}
    \label{fig:partcolouredmap}
\end{figure*}  

\begin{figure}[h!]
    \centering
    \subfloat[Popup showing the florets from user-selected polygons.]{%
        \includegraphics[width=0.4\textwidth]{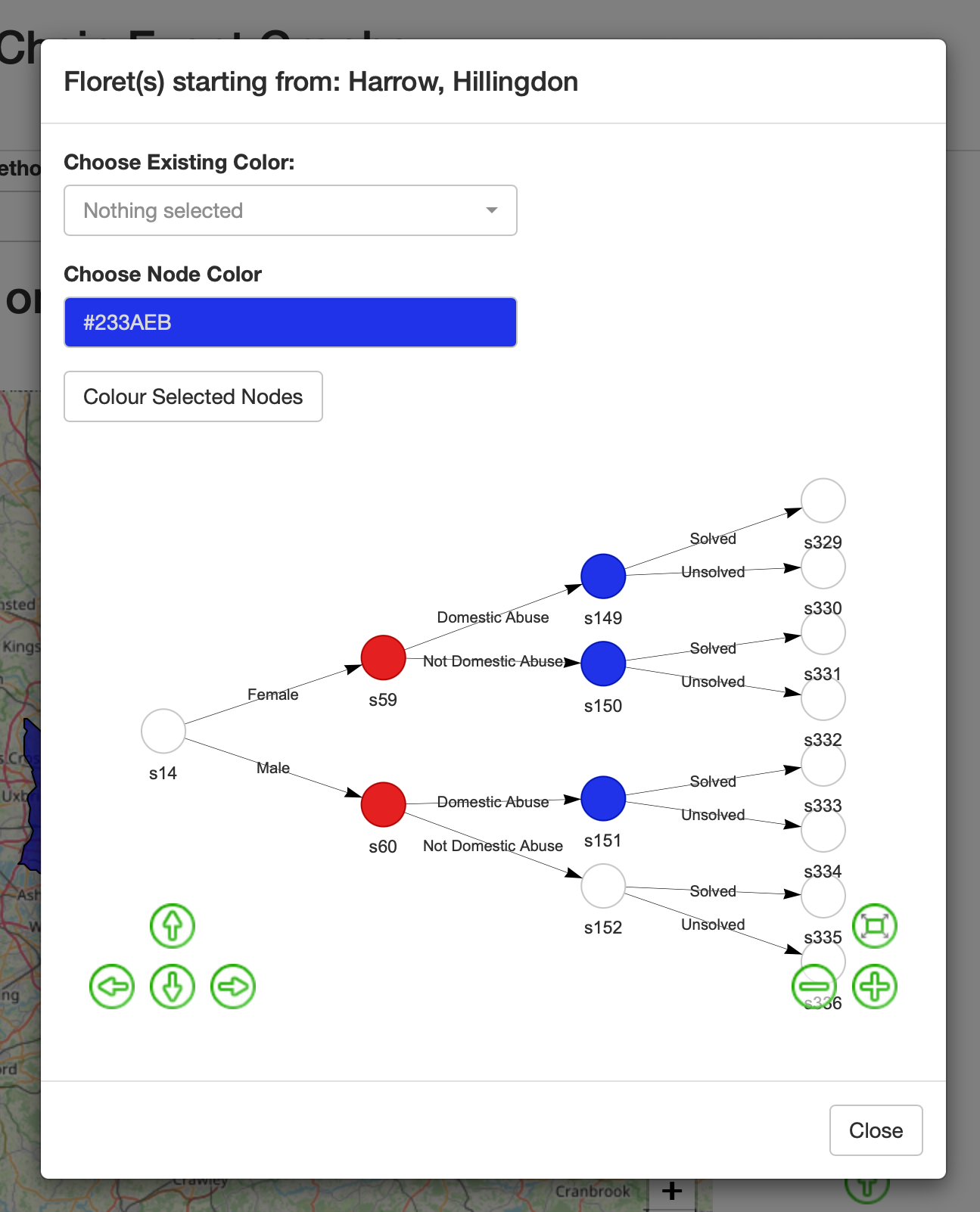}
        \label{fig:hillingdonfloret}
    }
    \subfloat[Event tree showing floret colourings.]{
    {\includegraphics[width=0.4\textwidth]{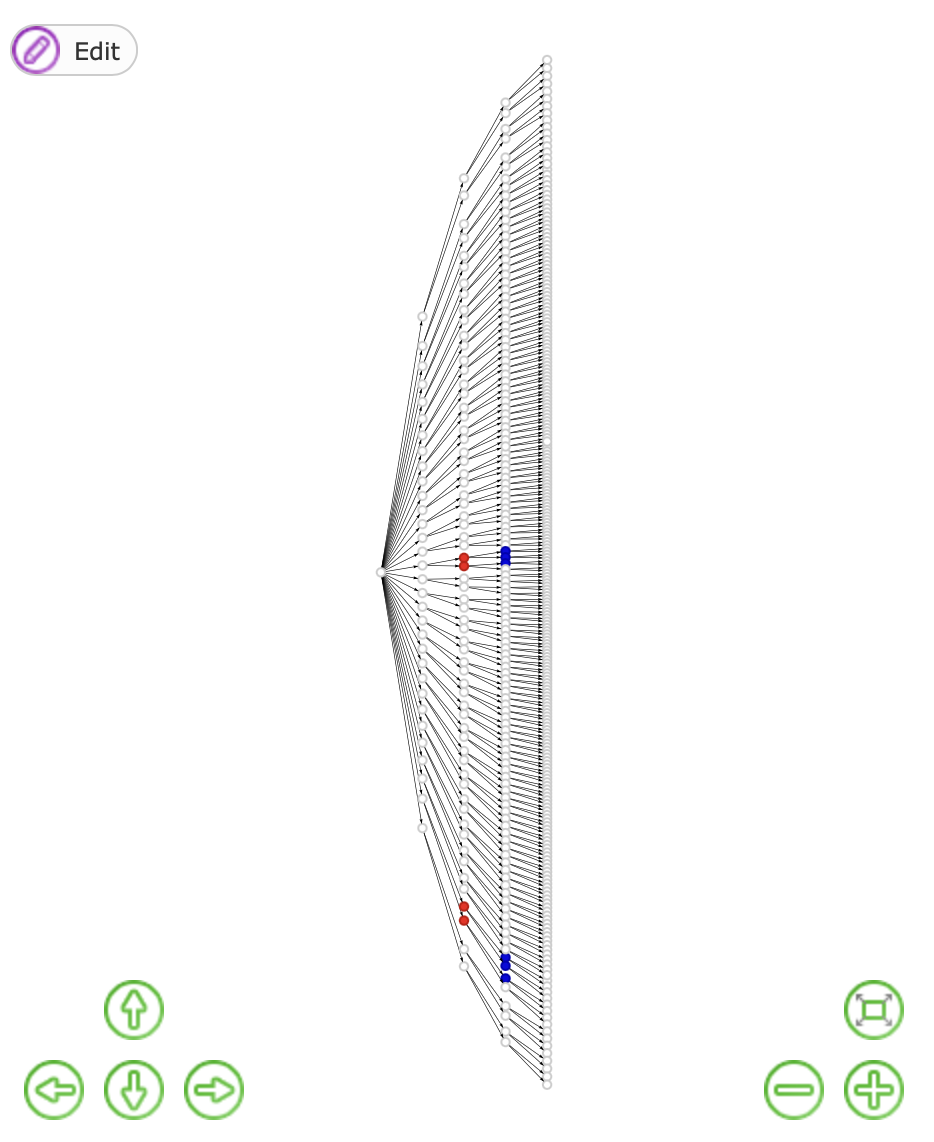}}
        \label{fig:hillingdontree}
    }
    \caption{Floret colouring pop-up and respective colourings present on the larger event tree. Note, as these florets are partially coloured, their respective polygons would be amber.}
    \label{fig:partialcolouring}
\end{figure}

\texttt{Map} is designed to be used when the event tree structure is large and contains an area variable (see Figure \ref{fig:partcolouredmap}). This allows the user to upload a zipped shapefile covering the areas in the tree. Any other polygons present will be ignored for colouring and rendered in grey. There is an option for the user to specify a Coordinate Reference System (CRS) for the file if one is not already present in order to ensure spatial data is interpreted correctly. Clikcing the \texttt{Process Shapefile} button renders a \textbf{leaflet} map, with the ability to adjust the layer opacity using the slider to see the terrain underneath. Initally, all polygons are displayed as dark orange, indicating that their area florets have not been coloured yet. By selecting polygons (highlighted in blue), and then clicking \texttt{Show Floret}, a pop-up appears, showing the floret for the area(s) (see Figure \ref{fig:hillingdonfloret}). This floret is then coloured the same way as in the larger event tree, by clicking on the nodes and applying the chosen colour. The user can also choose an existing colour already present in the larger tree, using the drop-down option, if they wish to add a node to an already existing stage. Once the pop-up is closed, this colouring will appear on the larger event tree for the respective areas (Figure \ref{fig:hillingdontree}). If the floret is fully coloured, the representative polygons appear green on the map, whereas if it is partially coloured (some situations remain white), they will appear amber. As in Section \ref{subsec:colouringtree}, the \texttt{Colour using AHC} button (found at the bottom of the sidebar) can either be applied to any remaining uncoloured nodes, or applied initially and then individual florets can be edited. When all polygons are green, representing that the tree has been fully coloured, the user can click the \texttt{Finished Colouring} button, saving the colouring. 
\vspace{-1.2em}

\textbf{Step 5}:
Clicking the \texttt{Finished Colouring} button creates a corresponding table displaying the stages. The user can then choose from the same prior options as for the \texttt{specify$\_$priors} function described in Section \ref{subsec:specify_priors} (\texttt{Custom}, \texttt{Uniform}, \texttt{Phantom Individuals}). If either of the latter two options are chosen, the user can make edits to certain stages if desired by clicking on the corresponding entries in the table. Once these priors have been specified for all stages, the user must click \texttt{Finished Prior Specification} to proceed.

\begin{figure*}[h!]
    \centering
    \makebox[\textwidth]{ 
    \includegraphics[width=0.9\textwidth]{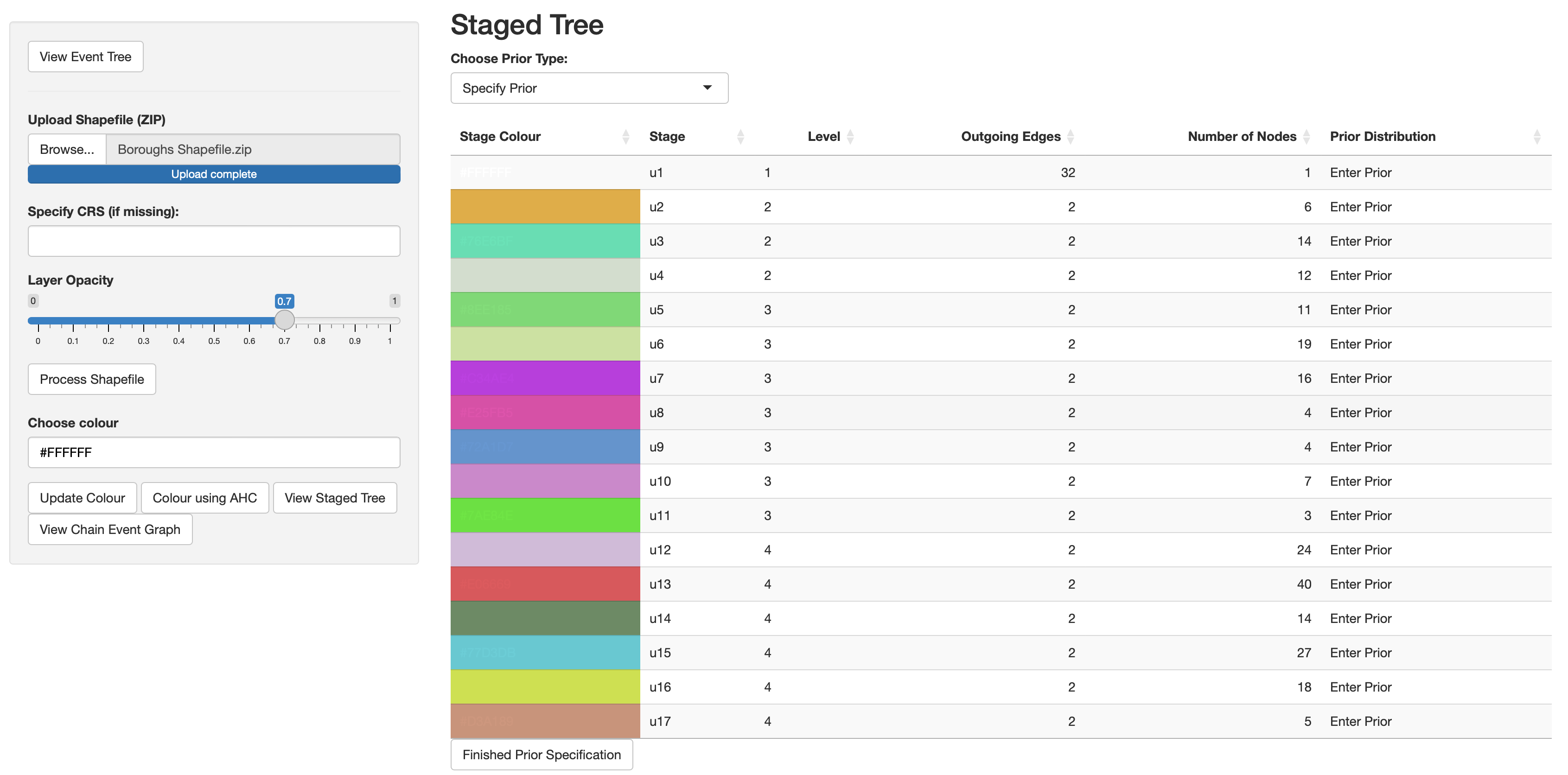}}
    \caption{The \texttt{Plots} panel in the \textbf{stCEG} web application, showing how the user can specify the priors for their staged tree.}
    \label{fig:shinypriors}
\end{figure*}
\begin{figure*}[h!]
    \centering
    \makebox[\textwidth]{ 
    \includegraphics[width=0.96\textwidth]{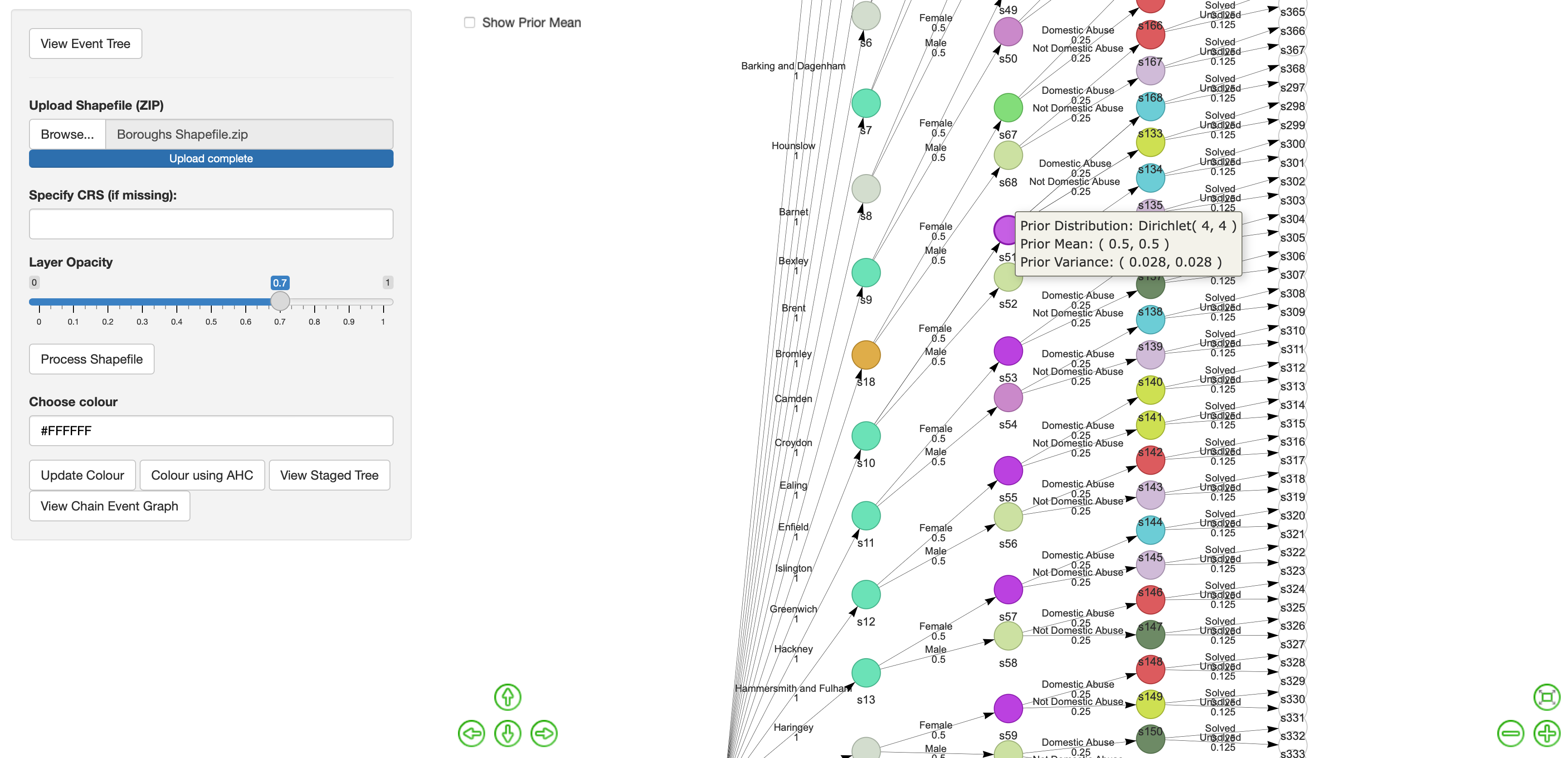}}
    \caption{The staged tree created in the \texttt{Plots} panel, given the user-entered priors. Hovering over a node displays the stage parameters, prior means and prior variances.}
    \label{fig:shinystagedtreehover}
\end{figure*}

   
\vspace{-0.5em}
These priors can then be visualised on the tree by clicking \texttt{View Staged Tree}. Using the tickbox above the plot, the user can choose to either view the Dirichlet parameters for each situation, or the prior mean. Hovering on a node gives a pop-up showing the stage parameters, prior means and prior variances. Figure \ref{fig:shinystagedtreehover} shows an example of this. 

\begin{figure*}[ht!]
    \centering
    \makebox[\textwidth]{ 
    \includegraphics[width=0.96\textwidth]{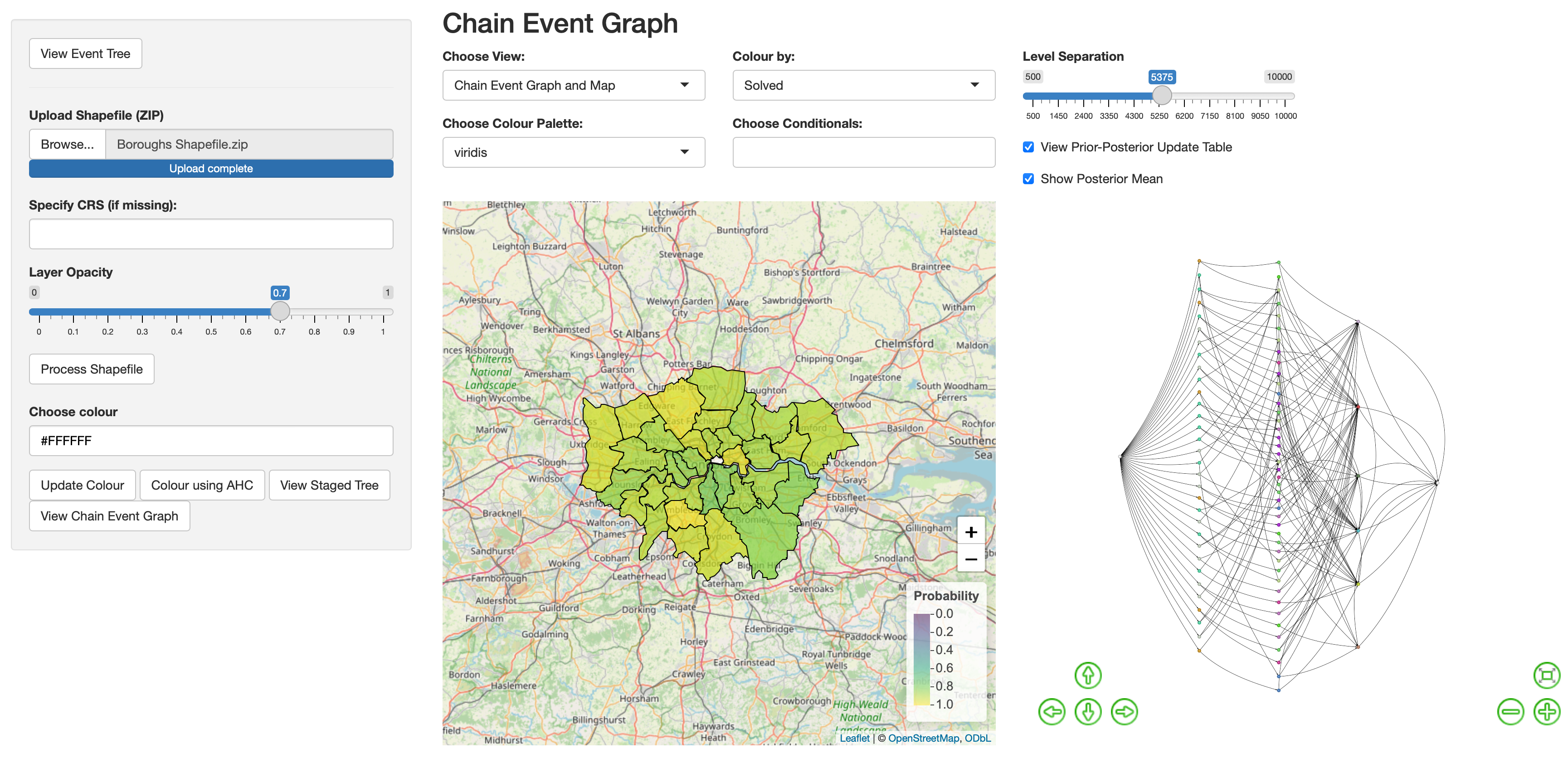}}
    \caption{The CEG created in the \texttt{Plots} panel. Clicking on a node colours its outgoing and incoming edges. Nodes can be drageed to reorder them.}
    \label{fig:shinyCEG_Map}
\end{figure*}

\textbf{Step 6}:
The \texttt{View Chain Event Graph} button contracts the nodes that share the same colour and structure to create the corresponding CEG (Figure \ref{fig:shinyCEG_Map}). The user can interact with this plot by dragging nodes to reorder them, and altering the space between levels using the \texttt{Level Separation} slider. To increase readability, the edges entering the selected node are coloured blue, and those emanating are coloured red.

\vspace{-0.5em}
As with colouring the event tree, users have the option to display either the CEG alone or alongside a map when a shapefile is available. In cases where a spatial variable has been included, the model structure can often become large, making it harder to elicit from. The addition of a map assists with this in 2 ways:
\vspace{-0.5em}
\begin{enumerate}
    \item The user can click an area on the map and this gives a popup with the reduced CEG for that area. 
    \begin{figure*}[h!]
    \centering
    \makebox[\textwidth]{ 
    \includegraphics[width=0.42\textwidth]{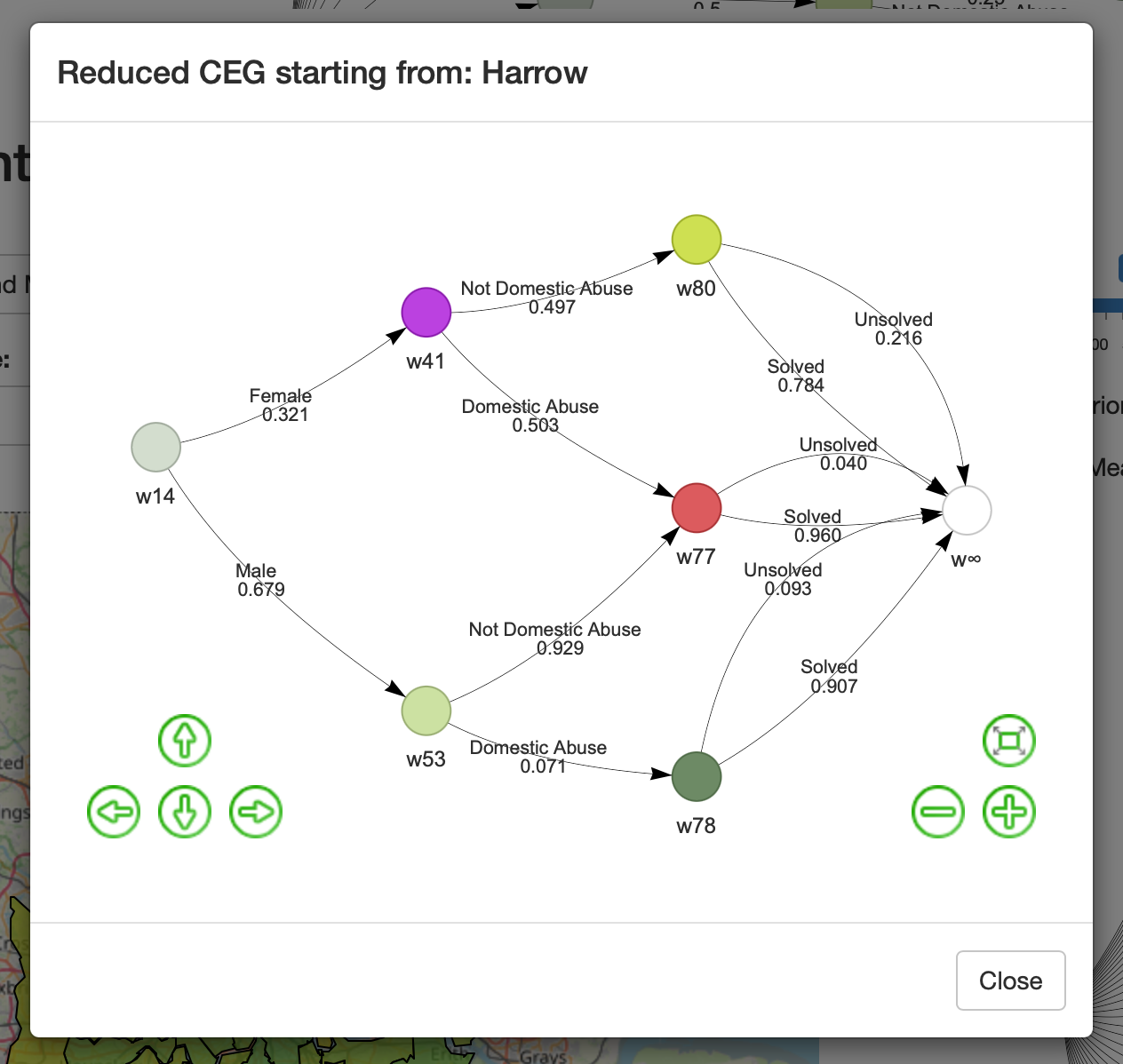}}
    \caption{The pop-up showing the reduced CEG for Harrow.}
    \label{fig:ShinyReducedCEG}
\end{figure*} 

    \item The map also allows the user to easily see the results of the model across each area. The user can change the colour palette and the variable it is coloured by using the drop-down boxes. Another useful feature is the ability to condition on the model to produce maps for different subsets of factors (e.g. Female, Domestic Abuse). This allows the user to reduce the model in a different way in order to perform categorical comparisons. Examples of this can be found in Figure \ref{fig:mapcolouring}. Conditioning on ``Male'' and ``Domestic Abuse'' shows us that there are four boroughs experiencing low solved probabilities, whereas these are not apparent on the map conditioned on ``Female'' victims.

    \begin{figure}[ht!]
    \centering
    \subfloat[Map coloured by \texttt{Solved}, conditioned on \texttt{Female}.]{%
        \includegraphics[width=0.42\textwidth]{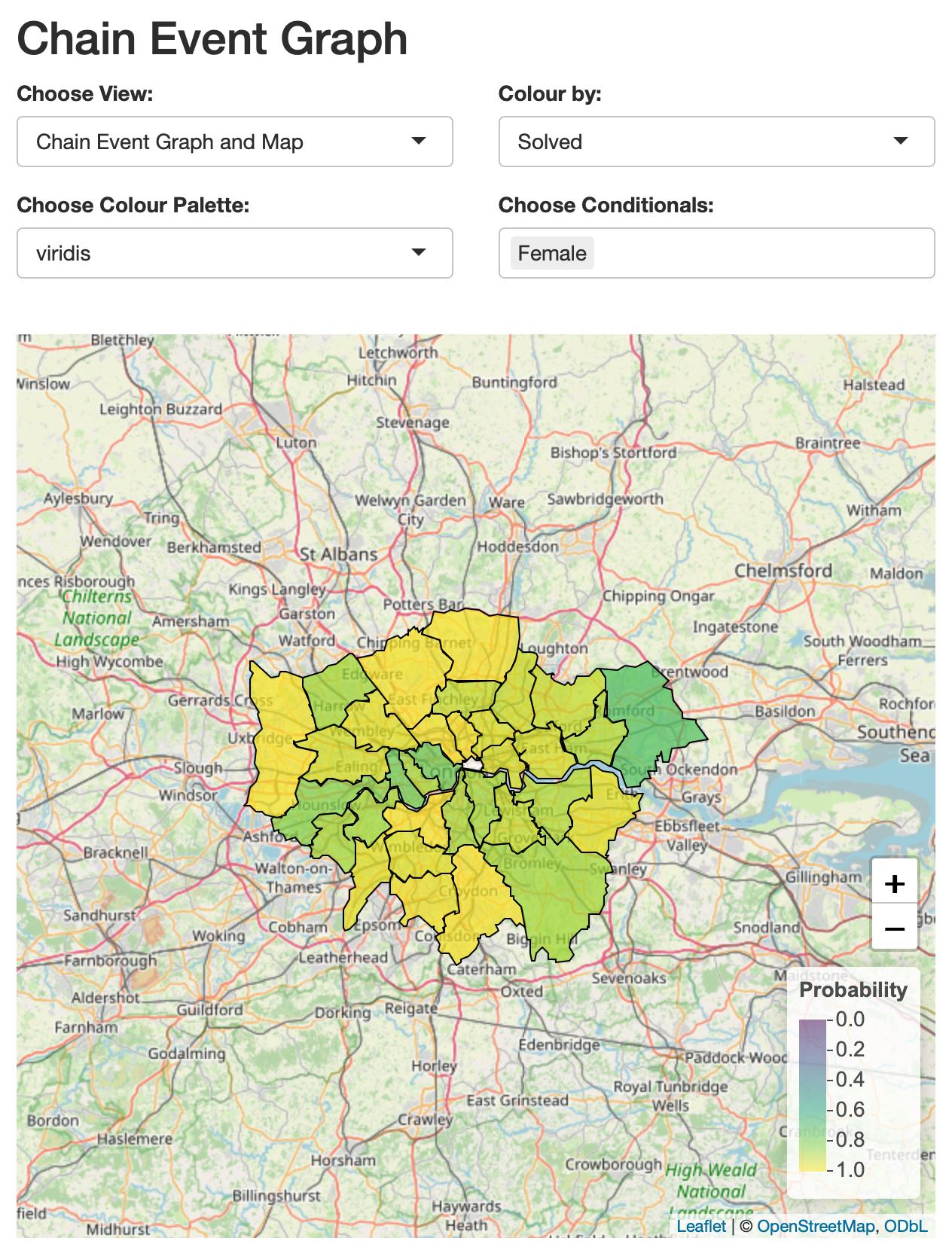}
        \label{fig:femaleconditional}
    }
    \hspace{1.5em}
    \subfloat[Map coloured by \texttt{Solved}, conditioned on \texttt{Male} and \texttt{Domestic Abuse}.]{
    {\includegraphics[width=0.42\textwidth]{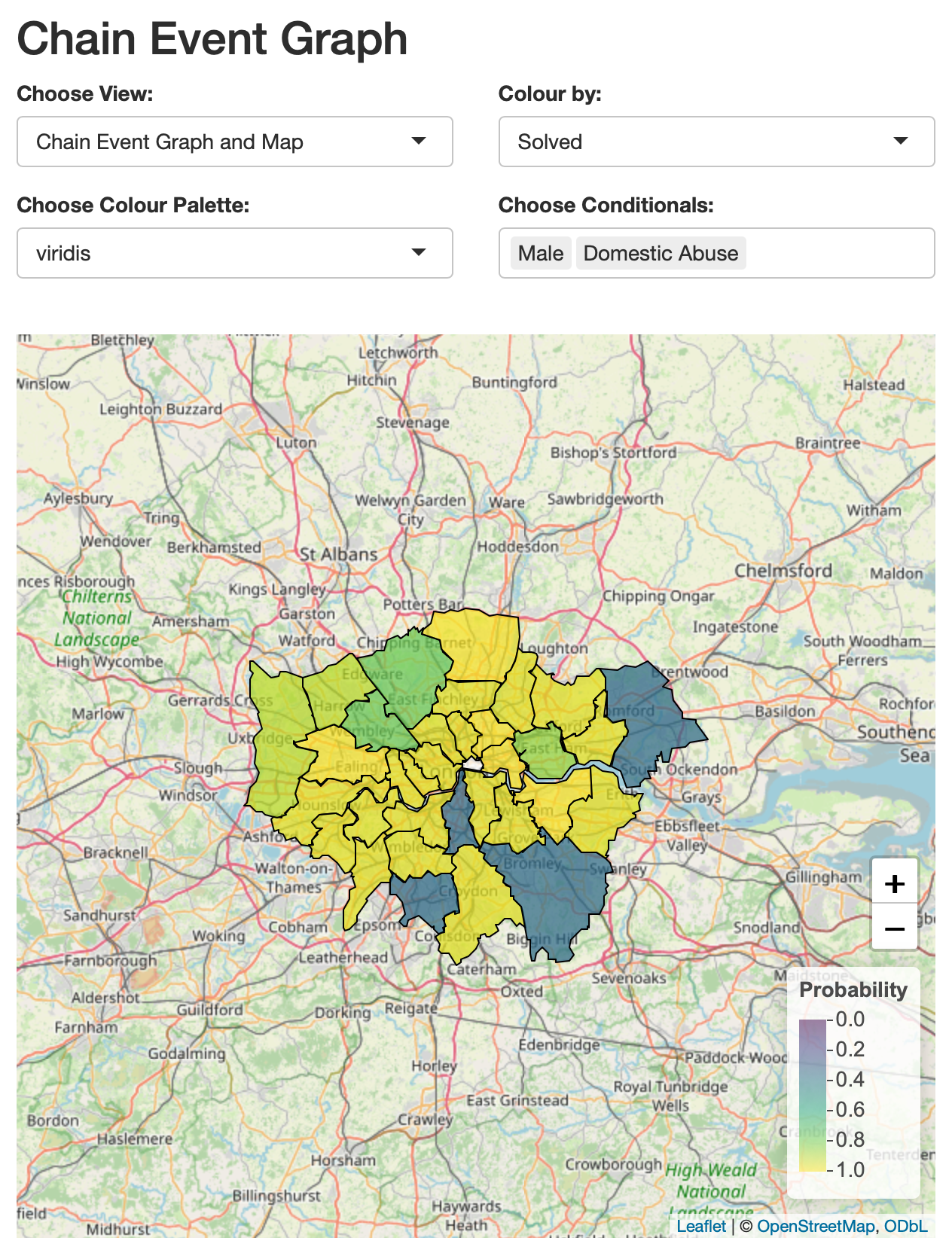}}
        \label{fig:maledaconditional}
    }
    \caption{Conditional maps for the CEG model in Figure \ref{fig:shinyCEG_Map} for a) Female, b) Male and Domestic Abuse.}
    \label{fig:mapcolouring}
\end{figure}
\end{enumerate}
The user may wish to see the Dirichlet updates for each stage. This could be to see how stages have evolved through the addition of data, or to see the impact of the specified priors on the final distribution. Figure \ref{fig:shinyCEG_Map} shows a tickbox under the \texttt{Level Separation} slider, which when ticked, generates this information in a table with the same information as in Figure \ref{tab:stprior}.  An example of this table output in \textbf{stCEG} can be seen in Appendix \ref{stcegshinytable}.

\section{Discussion}

\textbf{stCEG} is an open-source package in R for modelling staged trees and CEGs with both symmetric and asymmetric structures. This is the first package that allows the user to fully specify a CEG model, both in colouring and parameters. Moreover, it is uniquely able to handle spatial variables, with visualisations optimised for this. The following discusses several directions in which the package could be further developed.

We have presented in this package one implementation of how space can be encoded in a CEG, however there are limits to this approach. By simply treating space as another categorical variable, we are unable to encode that areas in close proximity may affect each other. Future research aims to extend the spatial theory surrounding CEGs in order to come up with a method that would allow dependence between different polygons whilst still retaining model interpretability.

To leverage the existing ability within \textbf{stCEG} to fully specify a model colouring and parameters, a development could be adding functionality to allow the user to specify an event tree structure themselves, without the use of a dataframe. This would enable the user to further explore the encoding of expert judgement, in the exploration of hypothetical scenarios, comparisons between alternative tree structures, or inclusions of theoretical constraints. 

Another direction for development lies in extending the temporal modelling capabilities of the package. Whilst the current version supports staged trees with spatial asymmetry, combining these with time-evolving processes would offer a richer class of dynamic models. This would involve adapting current temporal CEG methods to account for spatially-indexed variables, enabling users to explore how staged dependencies evolve over both time and space.

Finally, further efforts could be made to integrate model selection techniques and different scoring criteria within \textbf{stCEG}. This would be particularly valuable in contexts where multiple plausible event tree structures or stage colourings exist, reflecting alternative expert interpretations or theoretical assumptions. Users could weigh competing models not only by statistical fit but also by their practical consequences. This would strengthen the role of CEGs as tools for transparent evidence-based decision making.


\section*{Computational details}

The results in this paper were obtained using
\textbf{R}~4.4.1 with the user interface and supporting visualisations developed using the
\textbf{shiny}~1.10.0 and \textbf{leaflet}~2.2.2 packages. R itself
and all packages used are available from the Comprehensive
R Archive Network (CRAN) at
\url{https://CRAN.R-project.org/}. \textbf{stCEG} is also available from GitHub (\url{https://github.com/holliecalley/stCEG}).

\section*{Acknowledgments}

\begin{leftbar}
HC was supported by the Engineering and Physical Sciences Research Council (EPSRC) DTP Studentship EP/W524451/1. DW was supported by AI for Net Zero grant EP/Y005597/1: ADD-TREES. For the purpose of open access, the author has applied
a Creative Commons Attribution (CC BY) licence to any Author Accepted Manuscript
version arising from this submission. The research data supporting this publication are provided within the \textbf{stCEG} package.
\end{leftbar}

\newpage


\bibliographystyle{apalike}  
\bibliography{references}


\newpage
\begin{appendix}
\section{Dirichlet Prior to Posterior Update} \label{app:dirichletcalculation}

We denote the conditional probability vector for each vertex in stage $u_i$ with $k_i$ outgoing edges by $\bm{\theta}_i = (\theta_{i1}, \theta_{i2},\ldots,\theta_{ik_i})$, where $\theta_{ij}$ is the probability that an individual at some vertex $v \in u_i$ travels along its \textit{j}th outgoing edge, where $j \in \{1,\ldots,k_i\}$. Both Collazo and Freeman argue for the use of the Dirichlet distribution for prior specification of CEGs due to Dirichlet-Multinomial conjugacy, following on from the proof that Dirichlet priors eliminate intractability in Bayesian Networks by Geiger and Heckerman (\cite{Collazo2018ChainGraphs, Freeman2010LearningGraphs, Geiger1997AIndependence}).

The exchangeability assumption that all vertices in the same stage share the same probability distribution implies that the conditional probability vector $\bm{\theta}_i$ for stage $u_i$ with $k_i$ outgoing edges has a Dirichlet distribution $\text{Dir}(\bm{\alpha}_i) = \text{Dir}(\alpha_{i1},\ldots,\alpha_{ik_i})$. The joint density function of the prior is 
\begin{equation}
    \pi(\bm{\theta}_i) = \frac{\Gamma(\sum_{j=1}^{k_i}\alpha_{ij})}{\prod_{j=1}^{k_i}\Gamma(\alpha_{ij})}\prod_{j=1}^{k_i}\theta_{ij}^{\alpha_{ij}-1}
\end{equation}

\begin{equation}
     \pi(\bm{\theta}_i)\propto\prod_{j=1}^{k_i}\theta_{ij}^{\alpha_{ij}-1}   
\end{equation}

A data vector $\bm{y}$, where $\bm{y}_i$ is a vector of individuals passing through stage $u_i$ in the Chain Event Graph gives rise to the likelihood $p(\bm{y}_i|\bm{\theta}_i)$, defined by the Multinomial distribution

\begin{equation}
   p(\bm{y}_i|\bm{\theta}_i) \propto \prod_{j=1}^{k_i}\theta_{ij}^{y_{ij}}
\end{equation}

Using Bayes' Theorem, the joint density function of the posterior is

\vspace{-3mm}
\begin{equation} 
\begin{split}
    \pi(\bm{\theta}_i|\bm{y}_i) &\propto p(\bm{y}_i|\bm{\theta}_i)\pi(\bm{\theta}_i)\\
    &\propto \prod_{j=1}^{k_i}\theta_{ij}^{y_{ij}}\prod_{j=1}^{k_i}\theta_{ij}^{\alpha_{ij}-1}\\
 &\propto \prod_{j=1}^{k_i}\theta_{ij}^{\alpha_{ij}+y_{ij}-1}\\
 &\propto 
 \prod_{j=1}^{k_i}\theta_{ij}^{\alpha^*_{ij}-1}
\end{split}
\end{equation}
corresponding to a Dir($\bm{\alpha}^*_i$) = Dir$({\alpha^*_{i1}},\ldots,{\alpha^*_{ik_i}})$ distribution, where $\alpha^*_{ij} = \alpha_{ij} + y_{ij}$.

\newpage 

\section{AHC algorithm}
\label{sec:ahcalgorithm}
The Agglomerative Hierarchical Clustering algorithm is a greedy bottom-up search. In the context of CEGs, we start from an event tree where initially each situation is given its own stage.

We have a tree with stages $u_1,\ldots,u_k$, 
each stage having a conditional probability vector $\bm{\theta}_i$ and Dirichlet prior $\bm{\alpha}_i$. Given a data vector $\bm{y}_i$ of counts for each stage, we can calculate our posterior $\bm{\alpha}^*_i$, as described in Appendix \ref{app:dirichletcalculation}.

We compare each pair of stages $\{u_i, u_j\}$ with identical edge labels by computing the log Bayes Factor of the staged tree structure, $\mathcal{S'}$, that merges those situations into a single stage $u_{i\oplus j}$, compared to the structure, $\mathcal{S}$, that doesn't.

\begin{equation*}
    \text{log}BF(\mathcal{S'},\mathcal{S}) = \mathcal{L}(\mathcal{S'}) - \mathcal{L}(\mathcal{S})
\end{equation*}

where $\mathcal{L}(S)$ is the log-marginal likelihood of $\mathcal{S}$:

\begin{equation*}
    \mathcal{L}(S) = \sum_{i=1}^{k} \left[ 
    \psi\left( \bar{\alpha}_i \right) - \psi\left( \bar{\alpha}_i^* \right) 
    + \sum_{j=1}^{k_i} \left( \psi\left( \alpha_{ij}^* \right) - \psi\left( \alpha_{ij} \right) \right) 
\right]
\end{equation*}

with
\begin{itemize}
  \item $k$: Number of stages in the tree.
  \item $k_i$: Number of outgoing edges from stage $u_i$.
  \item $\alpha_{ij}$: Prior count for the $j$-th edge from stage $u_i$.
  \item $\alpha_{ij}^* = \alpha_{ij} + y_{ij}$: Posterior count for the $j$-th edge from stage $u_i$.
  \item $\bar{\alpha}_i = \sum_{j=1}^{k_i} \alpha_{ij}$: Total prior count for stage $u_i$.
  \item $\bar{\alpha}_i^* = \sum_{j=1}^{k_i} \alpha_{ij}^*$: Total posterior count for stage $u_i$.
  \item $\psi(x) = \log \Gamma(x)$
\end{itemize}

Expanding this we get:
\begin{equation*}
\begin{split}
\log BF(S', S) =\ 
&\psi\left( \bar{\alpha}_{i \oplus j} \right)
- \psi\left( \bar{\alpha}_i \right)
- \psi\left( \bar{\alpha}_j \right)
- \psi\left( \bar{\alpha}_{i \oplus j}^* \right) + \psi\left( \bar{\alpha}_i^* \right)
+ \psi\left( \bar{\alpha}_j^* \right) \\
&+ \sum_{l=1}^{k} \left[
    \psi\left( \alpha_{i \oplus j, l}^* \right)
    - \psi\left( \alpha_{i l}^* \right)
    - \psi\left( \alpha_{j l}^* \right)
    - \psi\left( \alpha_{i \oplus j, l} \right)
    + \psi\left( \alpha_{i l} \right)
    + \psi\left( \alpha_{j l} \right)
\right]
\end{split}
\end{equation*}

If there is no valid pair, we stop. Otherwise, we merge the pair with the highest logBF, summing the logBF values for the merged stages to form the CEG score and continue comparisons with the new structure.

When there are no more valid merges to make, we output the CEG score along with the final staged tree structure.
\newpage

\section{Code to Generate Comparison CEG}
\label{app:comparisoncode}
\begin{minipage}{\textwidth}
\textcolor{ForestGreen}{\texttt{\#Creating event tree}}\\
\texttt{homicides$\_$ET <- create$\_$event$\_$tree(dataset = homicides, columns = c(\textcolor{blue}{3},\textcolor{blue}{2},\textcolor{blue}{4},\textcolor{blue}{5})))}\\

\textcolor{ForestGreen}{\texttt{\#Colouring fully using AHC algorithm}}\\
\texttt{homicides$\_$AHC <- ahc$\_$colouring(homicides$\_$ET)}\\

\texttt{\textcolor{ForestGreen}{\#Uninformative Uniform priors}\\
priors$\_$AHC <- specify$\_$priors(homicides$\_$AHC,  \textcolor{ForestGreen}{"Uniform"}, ask$\_$edit = \textcolor{Periwinkle}{FALSE})}\\

\textcolor{ForestGreen}{\texttt{\#Adding priors to tree}}\\
\texttt{homicides$\_$ST$\_$AHC <- staged$\_$tree$\_$prior(homicides$\_$AHC, priors$\_$AHC)}\\

\textcolor{ForestGreen}{\texttt{\#Creating CEG}}\\
\texttt{homicides$\_$CEG$\_$AHC <- create$\_$ceg(homicides$\_$ST$\_$AHC, label = \textcolor{ForestGreen}{"posterior$\_$mean"})}\\
\end{minipage}

\section{Code to Generate Spatial CEG} \label{app:spatialcode}
\begin{minipage}{\textwidth}

\textcolor{ForestGreen}{\texttt{\#Takes a while to run as the model is large}}

\textcolor{ForestGreen}{\texttt{\#Creating event tree}}\\
\texttt{homicides$\_$ET$\_$spatial <- create$\_$event$\_$tree(dataset = homicides, columns = c(\textcolor{blue}{6},\textcolor{blue}{3},\textcolor{blue}{2},\textcolor{blue}{4},\textcolor{blue}{5}) , level$\_$separation = \textcolor{blue}{1300}, node$\_$distance = \textcolor{blue}{300}, label$\_$type = \textcolor{ForestGreen}{"both"})}\\

\textcolor{ForestGreen}{\texttt{\#Colouring fully using AHC algorithm}}\\
\texttt{homicides$\_$AHC$\_$spatial <- ahc$\_$colouring(homicides$\_$ET$\_$spatial, level$\_$separation = \textcolor{blue}{1300}, node$\_$distance = \textcolor{blue}{300})}\\

\textcolor{ForestGreen}{\texttt{\#Uninformative priors}}\\
\texttt{priors <- specify$\_$priors(homicides$\_$AHC$\_$spatial, \textcolor{ForestGreen}{"Phantom"}, ask$\_$edit = \textcolor{Periwinkle}{FALSE})}\\

\textcolor{ForestGreen}{\texttt{\#Adding priors to tree}}\\
\texttt{homicides$\_$ST$\_$spatial <- staged$\_$tree$\_$prior(homicides$\_$AHC$\_$spatial, priors, level$\_$separation = \textcolor{blue}{1300}, node$\_$distance = \textcolor{blue}{300})}\\

\textcolor{ForestGreen}{\texttt{\#Creating CEG}}\\
\texttt{homicides$\_$CEG$\_$spatial <- create$\_$ceg(homicides$\_$ST$\_$spatial, view$\_$table = \textcolor{Periwinkle}{TRUE}, label = \textcolor{ForestGreen}{"posterior$\_$mean"}, level$\_$separation = \textcolor{blue}{1500})}
\end{minipage}

\newpage

\section{stCEG Prior to Posterior Update Table}
\vspace{-0.5em}
\label{stcegshinytable}
\begin{figure*}[ht!]
    \centering
    \makebox[\textwidth]{ 
    \includegraphics[width=0.95\textwidth]{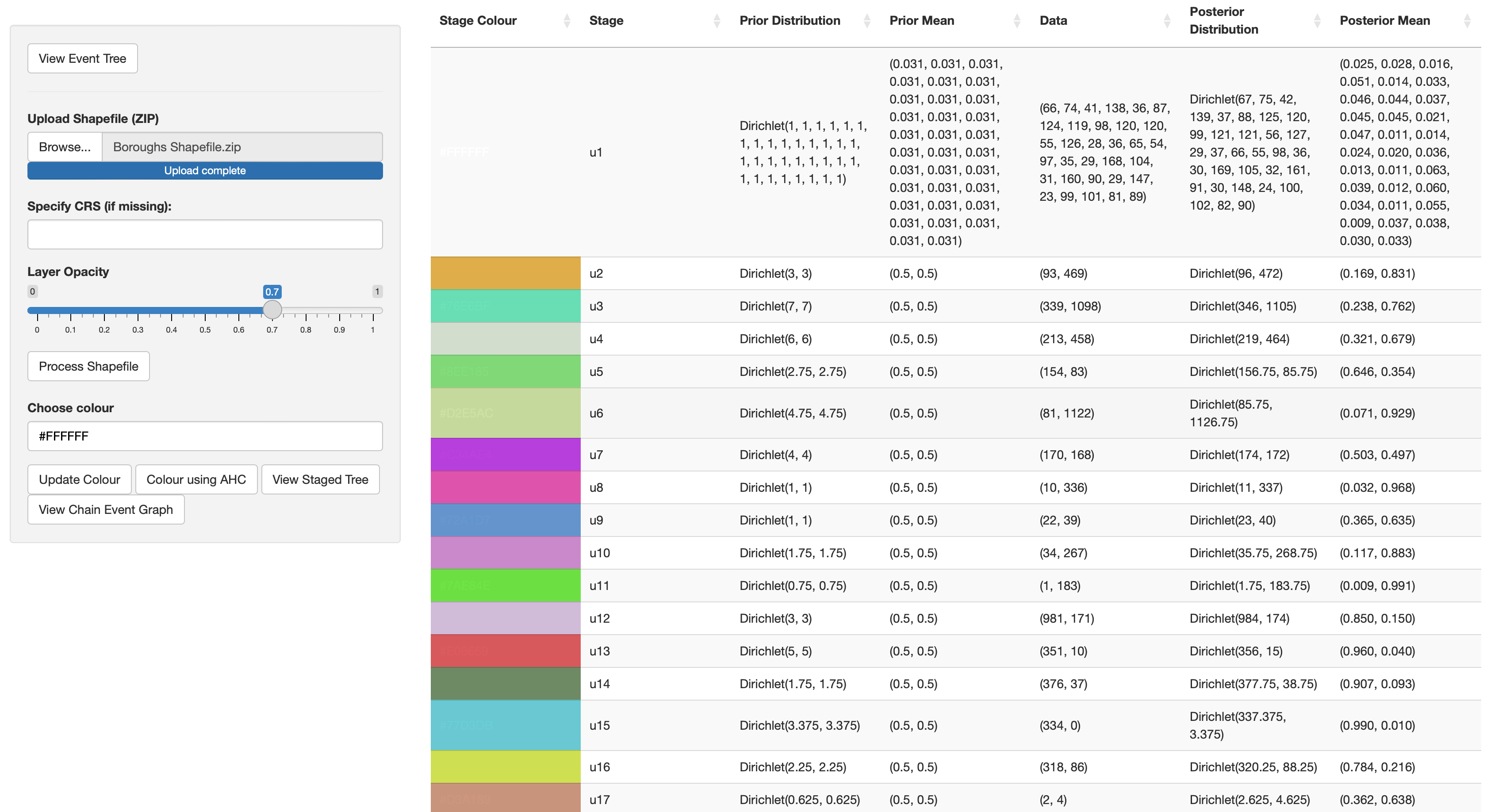}}
    \caption{The prior to posterior update table corresponding to the CEG in Figure \ref{fig:shinyCEG_Map}.}
    \label{fig:shinyposteriortable}
\end{figure*}

\section{stCEG UI Buttons}
\label{tablesshiny}

\begin{table}[h!]
\vspace{-1em}
\subsection*{Upload Data Tab:}
\vspace{0.5em}
\renewcommand{\arraystretch}{1.2} 
\begin{tabular}{|p{0.23\textwidth}|p{0.76\textwidth}|}
\hline
\texttt{Choose CSV File}                     & Clicking ``\texttt{Browse}'' will load local data files so the user can upload their desired dataset. \\ \hline
\texttt{Select Area Division}                & The user is able to choose an area division that can be used to filter the data in the next tab. If there are multiple different area columns in the data (e.g Local Authority Districts, LSOAs), then only one can be selected. If no area column is present in the data, the user need not fill this column in. \\ \hline
\texttt{Select Time Division}                & The user is able to choose a time division that can be used to filter the data in the next tab. At this point, there is no need to specify how the data is formatted. \\ \hline
\texttt{Exclude First Column as Row Numbers} & Checkbox indicating first column should be excluded. \\ \hline
\texttt{Header}                              & Checkbox indicating the first row of the data contains the column names. \\ \hline
\texttt{Separator}                           & Field separating character. \\ \hline
\texttt{Quote}                               & Quoting character. \\ \hline
\texttt{Display}                             & The user can choose between viewing the head of the data or all of the data. \\ \hline
\texttt{Select Prediction Variable}          & This is the variable that will be used as the final branch of the event tree/CEG. \\ \hline
\end{tabular}
\end{table}

\begin{table}[ht!]
\vspace{-1em}
\subsection*{Select Data Tab:}
\vspace{0.5em}
\renewcommand{\arraystretch}{1.2} 
\begin{tabular}{|p{0.23\textwidth}|p{0.76\textwidth}|}
\hline
\texttt{Choose Area}                & Given the user specified an area column in the previous tab, this option will appear for the user to select a subset of areas to be considered. If this option is left blank, all areas are considered. \\ \hline
\texttt{Select Time Type}           & Given the user specified a time column in the previous tab, they can select whether it is one of the following types: ``\texttt{Date}'', ``\texttt{Month-Year}'' or ``\texttt{Year}''. \\ \hline
\texttt{Specify Date Format}        & If the time type is either Date or Month-Year, the user is also required to specify the format of this date (e.g \%Y-\%m-\%d). \\ \hline

\end{tabular}
\end{table}

\begin{table}[ht!]
\renewcommand{\arraystretch}{1.2} 
\begin{tabular}{|p{0.23\textwidth}|p{0.76\textwidth}|}
\hline
\texttt{Select Date Range}          & A slider is dynamically rendered based on the previous two inputs, so the user can select a time range for the data to be filtered by. \\ \hline
\texttt{Choose Number of Variables} & The user can specify the number of variables that they want to include in the event tree. This does not include the prediction variable chosen on the previous tab, as that is automatically selected and placed at the end of the tree. \\ \hline
\texttt{Choose Variable} $x$        & Given a drop-down menu of the column names present in the dataframe, the user can choose however many variables they selected using the previous input. The order of these variables will match the order they are given in the tree. \\ \hline
\texttt{Set Default Selections}     & This randomly selects two areas (if an area column is present), and given a chosen number of variables, this button takes the variables in the order they appear in the dataframe. \\ \hline
\texttt{View Selection}             & This filters the data by the given inputs, and presents the filtered dataframe. \\ \hline
\end{tabular}
\label{tab:selectdatashiny}
\end{table}

\begin{table}[ht!]
\vspace{-0.5em}
\subsection*{Plots Tab:}
\vspace{0.5em}
\renewcommand{\arraystretch}{1.2} 
\begin{tabular}{|p{0.23\textwidth}|p{0.76\textwidth}|}
\hline
\texttt{View Event Tree}         & This creates an event tree based on the user selected variables and ordering.         \\
\hline
\texttt{Choose Colouring Method} & The two methods that can be used to colour the event tree are ``\texttt{Map}'' and ``\texttt{Event Tree}''. ``\texttt{Map}'' can only be used if one of the variables in the event tree is an area.                                                                                                                                                  \\
\hline
\texttt{Upload Shapefile}        & If the user wishes to colour using a map, then a corresponding zipped shapefile should be uploaded, with the polygon names matching the area names in the tree. The shapefile can contain polygons which aren't present in the tree, but these will be ignored for colouring.                                          \\
\hline
\texttt{Specify CRS}             & When working with shapefiles, sometimes a coordinate reference system (CRS) needs to be specified if it isn't stored in the information by default. A CRS is needed to ensure the shapefile aligns correctly on the map, especially when combining data from different sources or map systems. If the shapefile has loaded, but there is an error displaying it, potentially try inputting a CRS.                                                                           \\
\hline
\texttt{Layer Opacity}           & This slider dynamically adjusts the opacity of the coloured polygons displayed on the map.                                                                                                                                                                                                                             \\
\hline
\texttt{Process Shapefile}       & This loads the shapefile and renders a leaflet map with all polygons in the event tree coloured orange, representing that they have yet to be coloured.                                                                                                                                                                \\
\hline
\texttt{Show Floret}             & If at least one polygon on the map is selected, this button shows the corresponding floret from that area for colouring. If multiple polygons are selected, all associated florets will share colouring.                                                                                                               \\
\hline
\texttt{Choose Colour}           & By clicking either nodes in a floret popup, or nodes on the full event tree, the user can then choose a colour to colour them.                                                                                                                                                                                         \\
\hline
\texttt{Update Colour}           & This button takes the chosen colour and applies it to the selected nodes.                                                                                                                                                                                                                                              \\
\hline
\texttt{Colour using AHC}        & In the case that the user is uncertain about their judgements, they may choose to colour subsets of vertices using Agglomerative Hierarchical Clustering (see Section \ref{sec:stagedtreesandcegs}). If any vertices were previously coloured, these will retain their colouring. \\
\hline
\texttt{Delete Selected Node}    & In the case where the system of events has an asymmetric structure, the user can manipulate the tree to eliminate any nodes that are not needed by deleting them. This is analogous to the \texttt{delete$\_$nodes} function described in section \ref{subsec:createET}.                                                                            \\
\hline
\texttt{Finished Colouring}      & This finalises colour judgements so the user can specify their prior judgements for each of these stages. \\
\hline
\texttt{Choose Prior Type}         & Allows the user to select either ``\texttt{Specify Prior}'', ``\texttt{Uniform 1,1 Prior}'', or ``\texttt{Phantom Individuals Prior}'', and populates the stage table accordingly. If one of the pre-defined priors is chosen, the user can still edit individual priors manually in the table. \\ \hline
\texttt{Finished Prior Specification} & When the user is satisfied with the priors for each stage, this button is clicked to save them. \\ \hline
\texttt{View Staged Tree}          & Displays a staged tree with the same colouring as before. Prior parameters are shown as edge labels, and hovering over nodes reveals stage information. Clicking a checkbox toggles between showing edge labels as prior parameters or prior means. \\ \hline
\texttt{View Chain Event Graph}    & Creates the chain event graph (CEG) corresponding to the staged tree. Nodes are draggable, and clicking on a vertex colours incoming edges blue and outgoing edges red for easier readability. \\ \hline
\texttt{Level Separation}          & A slider that adjusts spacing between levels in the CEG. This is useful when many nodes or overlapping labels are present. \\ \hline
\end{tabular}
\end{table}

\begin{table}[ht!]
\renewcommand{\arraystretch}{1.2} 
\begin{tabular}
{|p{0.23\textwidth}|p{0.76\textwidth}|}
\hline
\texttt{Choose View}               & Lets the user choose whether to display a coloured map (if a shapefile was uploaded) alongside the CEG, or only the CEG. When a map is shown, clicking on a polygon produces a filtered CEG for that area. \\ \hline
\texttt{Choose Colour Palette}     & Dropdown menu for selecting the map's colour scale. Options include: \texttt{viridis}, \texttt{magma}, \texttt{turbo}, \texttt{plasma}, \texttt{inferno}, \texttt{cividis}, \texttt{mako}, and \texttt{rocket} (from \textbf{viridis} (\cite{Garnier2024Colorblind-Friendly0.6.5})). \\ \hline
\texttt{Colour By}                 & Lets the user select a value of the prediction variable (e.g.\ ``\texttt{Solved}'' or ``\texttt{Unsolved}'') to colour the map by. \\ \hline
\texttt{Choose Conditionals}       & A grouped dropdown menu by variable, allowing the user to condition the map on one or more selected values. This is analogous to the \texttt{conditionals} option in the \texttt{generate$\_$CEG$\_$map} function described in Section \ref{subsec:SpaceCEG}. \\ \hline
\end{tabular}
\vspace{52em}
\end{table}

\end{appendix}


\end{document}